\documentclass[10pt,aps,pra,showpacs,showkeys,twocolumn,superscriptaddress]{revtex4-1}
\usepackage[dvips]{color}
\usepackage{graphicx}
\usepackage{amssymb}
\usepackage{amsmath}
\begin{document}

\title{Entanglement dynamics of spins using a few complex trajectories}
\author{M. V. Scherer}
\author{A. D. Ribeiro}
\affiliation{Departamento de F\'{\i}sica, Universidade Federal do Paran\'a, 81531-990, Curitiba, PR, Brazil}
\date{\today}

\begin{abstract}

In this work, we consider two spins initially prepared in a product of coherent states and study their entanglement dynamics due to a general interacting Hamiltonian. We adopt an approach that allowed the derivation of a semiclassical formula for the linear entropy of the reduced density operator, assumed as an entanglement quantifier. The resulting expression depends on sets of four trajectories, originated from the underlying classical description, and having mutually connected final phase-space points. Such classical elements, which are capable to reproduce the quantum entanglement even for long values of propagation time, arise when we assume a proper analytical continuation of the classical phase space onto a complex domain. We apply this theory to a particular physical system, showing that taking into account only a few sets of complex trajectories is enough to get an excellent agreement between the semiclassical linear entropy of the reduced density operator and its quantum counterpart.


\end{abstract}


\maketitle

\section{Introduction}
\label{intro}

In 1935, Einstein, Podolsky, and Rosen~\cite{EPR} shook the structures of the quantum mechanics---which was just a newborn theory at that time---, asserting that the physical description provided by its formalism was not complete. The crisis generated by the authors had rapid answers given by the scientific community, among which we highlight an article written by Schr\"odinger~\cite{Sch1935}, where he coins the term {\em entanglement} and announces it as the phenomenon causing that surprising conclusion. In that contribution, Schr\"{o}dinger was incisive about the non-classical nature of entanglement, an idea that counterposes any attempt to readjust the quantum formalism with classical elements. 

Thirty years later, this discussion was re-expressed in solid mathematical grounds by Bell~\cite{bell}, who formulated the notion of local hidden variables, a construction that naturally emerges from the paper of Einstein and his collaborators~\cite{EPR}, and demonstrated that the quantum formalism cannot be supplemented by such a classical expedient. This result clearly favors Schr\"{o}dinger's argument expressing entanglement as~\cite{Sch1935} ``{\em the} characteristic trait of quantum mechanics, the one that enforces its entire departure from classical lines of thought.'' Another important point that should be mentioned is the fact that Bell's work has inaugurated the studies concerning the quantum statistical correlations, in which entanglement has a crucial role, culminating with the emergence of the quantum information theory~\cite{amicoRMP,horodecki,chuang}.

While a significant part of researchers has made efforts to unravel entanglement inside the quantum formalism, many others have been interested in understanding aspects of the quantum-classical connection involving this concept~\cite{kyoko, sanders, angelo04, escobar, jacquod1, angelo05, marcel2005, prosen05, matzkin2006, matzkin2006EPL, brumer2007, jacquod2, pra2010, matzkin2010, bonanca2011, matzkin2011, pra2012, casati, berenstein, neil2016, zheng2017, pattanayak2017, arlans, ghose2019, quach, pappalardi, smerzi}. Despite the previous arguments, which disconnect entanglement from the classical mechanics, notice that this question is still legitimate: For systems prepared in the semiclassical regime, where both quantum and classical theories can give accurate predictions of their behavior, it is expected that two different entanglement manifestations appearing in the quantum description have distinct symptoms in the respective classical treatment. In this scenario, it is remarkable that almost the totality of the aforementioned papers deals, to some extent, with quantum-chaos approaches. Very shortly, this kind of investigation aims to associate the regime of the underlying classical dynamics---regularity or chaos---with properties of the quantum entanglement. 

Were we extracting from all these papers~\cite{kyoko, sanders, angelo04, escobar, jacquod1, angelo05, marcel2005, prosen05, matzkin2006, matzkin2006EPL, brumer2007, jacquod2, pra2010, matzkin2010, bonanca2011, matzkin2011, pra2012, casati, berenstein, neil2016, zheng2017, pattanayak2017, arlans, ghose2019, quach, pappalardi, smerzi} the most typical result, it would be the statement that the rate of entanglement growth for a given initial quantum state, due to the action of an interacting Hamiltonian, is greater when its classical counterpart experiments a chaotic dynamics, compared with a regular regime. It is implied in this sentence that the quantum state should be initially well-localized in order to be eligible for a proper classical treatment. Meanwhile, we also need to point out that some works have questioned this characterization~\cite{matzkin2006, matzkin2006EPL, matzkin2010, matzkin2011, pattanayak2017, ghose2019}, showing that some states, whose classical counterparts do present regular dynamics, have entanglement growth which would be compatible with chaotic behavior.

In the present paper, we semiclassically investigate the entanglement dynamics between two spins, adding new elements in this research field. Our work can be seen as an extension of another contribution~\cite{arlans}, where only canonical degrees of freedom were considered. Here, we contemplate spin variables. The approach consists of taking the linear entropy of the time-dependent reduced density operator to characterize entanglement and deriving its approximated expression. The starting point for this calculation is the semiclassical forward and backward propagators in the spin-coherent-state representation. The resulting formula only depends on the trajectories originated from the underlying classical description of the problem. If we restrict the classical phase space to (ordinary) real coordinates, then the semiclassical entropy will be an exclusive function of the trajectory departing from the centroid of the initial state, assumed as a spin-coherent state. In this case, our result agrees with the most common conclusion seen in the literature, as it implies that entanglement behaves according to the inverse of the stability of that classical trajectory. This partial result was already published by one of us around ten years ago~\cite{pra2010}, and satisfactorily reproduces the exact calculation just for short evolution times. 

The novelty here, besides the treatment of spin variables, is analytically extending the real phase-space to the complex domain, so that we were able to identify new (complex) trajectories contributing to the semiclassical linear entropy. More precisely, we found that sets of four classical trajectories with final coordinates mutually connected are involved in the semiclassical computation. We also show, for a particular Hamiltonian, that the consideration of just a few sets of trajectories produces excellent results when compared with the quantum calculation. 

We organize the paper as follows. In Sect.~\ref{sle}, while important preliminary results are introduced---as the coherent-state representation (Sect.~\ref{scs}), the semiclassical forward and backward propagators (Sect.~\ref{sp}), and the saddle point method (Sect.~\ref{spm})---, we derive the formula for the semiclassical linear entropy. After that, in Sect.~\ref{ntv}, we make a first attempt to understand what is behind the new contributing trajectories, looking for them in the vicinity of real trajectories. At last, a numerical application of the theory is performed in Sect.~\ref{example1} and our final remarks are presented in Sect.~\ref{fr}.

\section{Semiclassical linear entropy}
\label{sle}

Taking $\hat\rho$ as a pure density operator representing a state consisting of two parts, $A$ and $B$, we can evaluate their entanglement by calculating the linear entropy of the reduced state $\hat\rho_A = \mathrm{Tr}_B(\hat\rho)$, expressed by
\begin{equation}
S(\hat\rho_A) = 1 - P(\hat\rho_A).
\label{Sdef}
\end{equation}
This quantity is symmetric, in the sense that $S(\hat\rho_A)=S(\hat\rho_B)$, and $P(\hat\rho_A)=\mathrm{Tr}_A[\hat\rho_A^2]$ is the purity of $\hat\rho_A$. Essentially, if the total state $\hat\rho$ is separable, then the purity of its reduced states is always equal to 1. On the other hand, for the case of entangled states, the purity lies in the range $0\le P(\hat\rho_A)<1$. Clearly, these properties promote $S(\hat\rho_A)$ to a kind of entanglement sensor for pure bipartite states: it returns 0, for separable states, and a value such that $0<S(\hat\rho_A)\le1$, for the entangled ones.

In the present paper, we are interested in describing entanglement as a function of time $T$, so that it is convenient to write the state under investigation as $\hat\rho_T=\hat{U}_T\hat\rho_0\hat{U}_T^\dagger$, where $\hat{U}_T$ is a general time-evolution operator. Concerning the initial state $\hat\rho_0$, one reminds that it must be considered as a pure state, that is, $\rho_0=|\psi_0\rangle\langle\psi_0|$. For the sake of clearness, we will initially consider a discrete basis $\{|n_A\rangle\otimes|n_B\rangle\}$, where $\{|n_{A}\rangle\}$ spans $\mathcal{H}_{A}$, the Hilbert space assigned to part $A$, and the same for part $B$. Given all these points, in order to derive a semiclassical formula for Eq.~(\ref{Sdef}), notice that we simply need to deal with 
\begin{equation}
\begin{aligned}
P_T=&\sum_{n_A,\ldots,m_B}
     \langle n_A,n_B|\hat{U}_T|\psi_0\rangle \langle \psi_0|
     \hat{U}_T^\dagger|m_A,n_B\rangle\\
    &\times 
     \langle m_A,m_B|\hat{U}_T|\psi_0\rangle \langle \psi_0|
     \hat{U}_T^\dagger|n_A,m_B\rangle,
\end{aligned}
\label{Pdiscrete}
\end{equation}
which is an explicit formula for the purity of a reduced state obtained from $|\psi_0\rangle$, evolved in time according to a general $\hat{U}_T$.

As our goal is studying spin systems in a semiclassical approach, the next step is rewriting Eq.~(\ref{Pdiscrete}) in terms of a more appropriate basis. We will adopt the spin-coherent-state representation, introduced in the following.

\subsection{Spin coherent states}
\label{scs}

For simplicity, we will begin the present discussion restricted to only one part of the bipartite system. Later on, results will be straightforwardly extended to the whole state. We also need to comment that there are several references dealing with this subject~\cite{radcliffe,klauderb,perelomov,gilmore,gazeau}, from which this subsection was written.

Essentially, a spin coherent state $|s\rangle$ is interpreted as the most classical spin state, in the sense that it saturates the uncertainty relation~\cite{ur} for the angular momentum operator $\hat{\mathbf{J}}=(\hat{J}_x,\hat{J}_y,\hat{J}_z)$,
\begin{equation}
\langle \Delta \hat{J}_{a}^2\rangle  
\langle \Delta \hat{J}_{b}^2\rangle \ge
\frac{1}{4}|\langle [\hat J_{a},\hat J_{b}]\rangle|^2+
\frac{1}{4}|\langle \{\Delta\hat J_{a},\Delta\hat J_{b}\}\rangle|^2,
\nonumber
\end{equation}
where $a$ and $b$ can assume $x$, $y$, or $z$. For the $(2j+1)$-dimensional case, $|s\rangle$ is explicitly given by
\begin{equation}
|s\rangle=
\frac{\exp{\left\{s\hat{J}_+\right\}}}
{\left(1+|s|^2\right)^{j}}
|-j\rangle,
\label{spin}
\end{equation}
where $\hat{J}_+ = \hat{J}_x+i\hat{J}_y$ is the raising spin operator and the ket~$|-j \rangle$ is the extremal eigenstate of $\hat{J}_z$ with eigenvalue~$-j$. Notice that $\hat{\mathbf{J}}$ and the complex number $s$, used to label the state, are dimensionless quantities.

An interesting and useful way to represent a coherent state can be reached from its mean value
\begin{equation}
\langle\hat{\mathbf{J}}\rangle=j\mathbf{n},
\nonumber
\end{equation}
where $\mathbf{n}$ is a unitary vector conveniently written in spherical coordinates as $\mathbf{n}=(\sin\theta \cos \phi,\sin\theta \sin \phi,\cos \theta)$. The connection between the label $s$ and $\mathbf{n}$ becomes particularly simple when a stereographical projection involving the complex plane $s$ and the unitary sphere is considered. Making the south pole correspond to $s=0$, while the north pole corresponds to $s\to\infty$, we find that 
\begin{equation}
s=|s|e^{-i\phi},\quad\mathrm{with}\quad
|s|=\cot (\theta/2).
\label{sphere}
\end{equation}
Since $|s\rangle$ are minimum-uncertainty states, it becomes natural to connect $\langle \hat{\mathbf{J}} \rangle$ with a classical angular momentum~$j\mathbf{n}$, which justifies the appeal to use them in semiclassical approaches. 

In practice, we can say that the wide use of coherent states in many research fields is somehow due to their capacity of generating a basis for the states in Hilbert's space. In particular, it allows for defining an over-complete unity resolution
\begin{equation}
\int |s \rangle \langle s | \, \mbox{d}\mu(s) = \hat{1},\quad
\mbox{d}\mu(s) \equiv \frac{2j+1}{\pi} \frac{\mbox{d}s^{{(R)}}\mbox{d}s^{{(I)}}}
{\left( 1 + |s |^2 \right)^{2}},
\label{NO1}
\end{equation}
where $s^{{(R)}}$ and $s^{{(I)}}$ are, respectively, the real and the imaginary part of $s$, and the integral runs from $-\infty$ to $+\infty$. In addition, it is important to mention that spin coherent states are non-orthogonal, 
\begin{equation}
\langle s | \tilde{s} \rangle = 
\frac{\left( 1 + s^* \tilde{s} \right)^{2j}}
{\left( 1 + |\tilde{s} |^2 \right)^{j}
\left( 1 + |s |^2 \right)^{j}} ,
\nonumber
\end{equation}
where $s^*$ ($\tilde{s}^*$) is the complex conjugate of $s$ ($\tilde{s}$).

We can now return to the original problem of entanglement in bipartite states, constraining the initial state as a product of two coherent states, that is, 
\begin{equation}
|\psi_0\rangle = |\mathbf{s}_0\rangle = 
|s_{0A}\rangle \otimes|s_{0B}\rangle \equiv |s_{0A},s_{0B}\rangle, 
\label{s0}
\end{equation}
where both $| s_{0A} \rangle$ and $| s_{0B} \rangle$ are given by Eq.~(\ref{spin}). Moreover, according to Eq.~(\ref{NO1}), we can rewrite Eq.~(\ref{Pdiscrete}) in the spin-coherent-state representation, finding   
\begin{equation}
\begin{aligned}
P_T=&\int\langle s_A,s_B| \hat{U}_T|\mathbf{s}_{0}\rangle \times
     \langle \mathbf{s}_{0}| \hat{U}_T^\dagger| \tilde{s}_{A},s_{B}\rangle\\
    &\times\langle\tilde{s}_A,\tilde{s}_B| \hat{U}_T|\mathbf{s}_{0}\rangle \times 
     \langle \mathbf{s}_{0}|\hat{U}_T^\dagger| s_{A},\tilde{s}_{B}\rangle ~
     \mbox{d}\boldsymbol{\mu},
\end{aligned}
\label{Pscs}
\end{equation}
where $\mbox{d}\boldsymbol{\mu}\equiv\mbox{d}\mu(s_A)\,\mbox{d}\mu(\tilde{s}_A)\,\mbox{d}\mu(s_B)\,\mbox{d}\mu(\tilde{s}_B)$. In the integrand, we identify four quantum propagators, defined by
\begin{equation}
\mathrm K_{\xi}\left( \mathbf s_{\eta},\mathbf s_{\mu}, T \right)=
\langle s_{\eta A},s_{\eta B}| e^{-i\xi\hat{H}T/\hbar} |s_{\mu A},s_{\mu B}\rangle,
\label{qp}
\end{equation}
where $\hat{H}$ is the Hamiltonian, $\hat{U}_T = e^{-i\hat{H}T/\hbar}$, $\mathbf{s}_{\eta}=(s_{\eta A},s_{\eta B})$, and $\mathbf{s}_{\mu}=(s_{\mu A},s_{\mu B})$. The label $\xi$ refers to forward ($\xi=+$) or backward ($\xi=-$) propagators.

Equation~(\ref{Pscs}) is proper to apply our semiclassical approximation, replacing each quantum propagator $\mathrm K_{\xi}$ with their respective semiclassical formulas $\mathcal{K}_\xi$. For this reason, in the next subsection, we briefly discuss such approximated expressions.

\subsection{Semiclassical propagator}
\label{sp}

There is a vast literature concerning the application of semiclassical approximations to the forward quantum propagator $\mathrm{K}_+$, for both canonical and spin degrees of freedom~\cite{scsp1,scsp2,scsp3,aguiar01,ribeiro04,garg1,sscsp1, sscsp2,sscsp3,sscsp4,sscsp5,ribeiro06,garg2,thiago}. On the other hand, concerning the backward propagator $\mathrm{K}_-$, we point out that we have worked with its semiclassical version in the last decade~\cite{pra2010,pra2012,foggiatto,arlans}. Very shortly, to deduce a semiclassical expression for $\mathrm{K}_\pm$, one starts from its path integral formulation, identifying, under proper assumptions ($j\to\infty$ and/or $\hbar\to0$), certain classical trajectories as the critical paths of integration. Thus, to conclude the approximation, the integrand is expanded up to the second order around them, and the resulting Gaussian integral is computed.

The classical trajectory involved in this kind of calculation has initial and final boundary conditions mandatorily related to the labels of the ket and the bra appearing in Eq.~(\ref{qp}). Such constraints are simplified when a new set of classical variables $\mathbf{u}=(u_A,u_B)$ and $\mathbf{v}=(v_A,v_B)$ is introduced, according to   
\begin{equation}
\tilde H(\mathbf u,\mathbf v) =
\tilde H(\mathbf s, \mathbf s^*)\equiv
\langle \mathbf s| \hat{H} | \mathbf s\rangle.
\label{Htilde}
\end{equation}
Using it, the mentioned boundary conditions become
\begin{equation}
\begin{array}{lll}
\mathbf u'= \mathbf s_\mu     &\quad \mathrm{and} \quad 
\mathbf v''=\mathbf s_\eta^*, &\quad\mathrm{for}\quad\xi=+,
\\
\mathbf u''= \mathbf s_\mu    &\quad \mathrm{and} \quad 
\mathbf v'=\mathbf s_\eta^*,  &\quad\mathrm{for}\quad\xi=-,
\end{array}
\label{bbnn}
\end{equation}
where (here and in the rest of the paper) we use the notation that a single (double) prime stands for initial (final) time. Classical equations of motion in the new variables are
\begin{equation}
\frac{\partial\tilde{H}}{\partial u_A}=
\frac{-2i\hbar j\dot{v}_A}{\left(1+u_Av_A\right)^2} 
\quad\mathrm{and}\quad
\frac{\partial\tilde{H}}{\partial v_A}=
\frac{2i\hbar j\dot{u}_A}{\left(1+u_Av_A\right)^2} ,
\label{emnn}
\end{equation}
and the equivalent for the variables assigned to part $B$.

The peculiarity of the semiclassical propagators in coherent states resides in the fact that the boundary conditions~(\ref{bbnn}) are generally overdetermined, provided that one keeps the natural assumption that $\mathbf{u}$ is the complex conjugate of $\mathbf{v}$, and vice-versa. In other words, Eq.~(\ref{bbnn}) implies the complete knowledge of both initial and final phase-space points. As a trajectory is determined by only one of them, the possibility of finding one that satisfies~(\ref{bbnn}) is reduced to the very singular case where $\mathbf{s}_\eta$ and $\mathbf{s}_\mu$ are fortuitously connected by the classical dynamics.

This problem is surpassed by realizing that the theory allows for the analytical continuation of the (real) classical phase space onto the complex domain. In this case, $\mathbf{u}$ and $\mathbf{v}$ are seen as independent variables and the searching for trajectories, which are now complex, becomes possible, in general. From this scenario, we emphasize the importance of distinguishing {\it real} and {\it complex} trajectories. While the former lives in the ordinary classical phase space, having the property $\mathbf{u}^*=\mathbf{v}$, for all instants of time, the last one inhabits an extended (complex) phase space and $\mathbf{u}^*\neq\mathbf{v}$. We should also comment that more than one trajectory satisfying Eqs.~(\ref{bbnn}) and~(\ref{emnn}) may exist, and all of them in principle should be used in the calculation of $\mathcal{K}_\xi$. 

The stability of the complex trajectories presented above is explicitly involved in the semiclassical calculations here studied. For this reason, it is important to define the stability matrix~$\mathbf{M}$ through the expression
\begin{equation}
\left(\begin{array}{c}
\delta \mathbf{u}''\\\delta \mathbf{v}''
\end{array}\right)=
\left(\begin{array}{cc}
\mathbf{M}_{\mathbf{uu}} & \mathbf{M}_{\mathbf{uv}}\\ 
\mathbf{M}_{\mathbf{vu}} & \mathbf{M}_{\mathbf{vv}}
\end{array}\right)
\left(\begin{array}{c}
\delta \mathbf{u}'\\\delta \mathbf{v}'
\end{array}\right) .
\label{stabmat}
\end{equation}
Essentially, $\mathbf{M}$ is responsible for the evolution of sufficiently small initial displacements $\delta\mathbf{u}'$ and $\delta\mathbf{v}'$ until the final time $T$. As we will show [see Eq.(\ref{StoM}) below], the elements of $\mathbf{M}$ can be written in terms of the second derivatives of the complex action $\mathcal{S}_\xi=\mathcal{S}_\xi(\mathbf s_\eta^*,\mathbf s_\mu,T)$, which is given by
\begin{equation}
\mathcal{S}_\xi =
\xi \int_0^T\left[ i\hbar j \chi-{\tilde H}\right]\mbox{d}t-
i\hbar j \tilde{\Lambda},
\label{S}
\end{equation}
where $\chi\equiv\chi_A+\chi_B$ and $\tilde{\Lambda}\equiv\tilde{\Lambda}_A+\tilde{\Lambda}_B$, with
\begin{equation}
\begin{aligned}
\chi_A\equiv            &\frac{\dot{u}_A v_A-u_A\dot{v}_A}{1+u_Av_A},\\
\tilde{\Lambda}_A\equiv &\ln\left[(1+u_A'v_A')(1+u_A''v_A'')\right],
\end{aligned}
\nonumber
\end{equation}
and the same for part $B$. By differentiating $\mathcal{S}_\xi$, we get the relations 
\begin{equation}
\begin{aligned}
\frac{\partial \mathcal{S}_+}{\partial u'_{A}} =
\frac{-2i\hbar jv_A'}{1+u_A'v_A'}, \qquad
&\frac{\partial \mathcal{S}_-}{\partial u''_{ A}} =
\frac{-2i\hbar jv_A''}{1+u_A''v_A''},\\
\frac{\partial \mathcal{S}_+}{\partial v''_{A}} = 
\frac{ -2i\hbar ju_A''}{1+u_A''v_A''}, \qquad
&\frac{\partial \mathcal{S}_-}{\partial v'_{A}} = 
\frac{ -2i\hbar jv_A'}{1+u_k'v_A'},
\end{aligned}
\label{variaS}
\end{equation}
and the analog for part $B$, which will be also very useful later. Now, if we differentiate Eq.~(\ref{variaS}) and properly rearrange the terms (for details, see Appendix~A of Ref.~\cite{pra2012}), we have 
\begin{equation}
\begin{array}{l}
\frac{i}{\hbar}\mathbf{S}^{(+)}_{\mathbf{u}'\mathbf{v}''} =
\left( \mathbf{A}' + \mathbf{B}' \right)\mathbf{M}^{-1}_{\mathbf{v}\mathbf{v}},\\
\frac{i}{\hbar}\mathbf{S}^{(-)}_{\mathbf{v}'\mathbf{u}''} =  
\left( \mathbf{A}' + \mathbf{B}' \right)\mathbf{M}^{-1}_{\mathbf{u}\mathbf{u}},\\
\frac{i}{\hbar}\mathbf{S}^{(+)}_{\mathbf{v}''\mathbf{v}''} = 
\left( \mathbf{A}'' + \mathbf{B}'' \right)\mathbf{M}_{\mathbf{u}\mathbf{v}}
\mathbf{M}^{-1}_{\mathbf{v}\mathbf{v}} - \mathbf{C}'',\\
\frac{i}{\hbar}\mathbf{S}^{(-)}_{\mathbf{u}''\mathbf{u}''} = 
\left( \mathbf{A}'' + \mathbf{B}'' \right)\mathbf{M}_{\mathbf{v}\mathbf{u}}
\mathbf{M}^{-1}_{\mathbf{u}\mathbf{u}} - \mathbf{D}'',
\end{array}
\label{StoM}
\end{equation}
where 
\begin{equation}
{\mathbf{S}}^{(\xi)}_{\mathbf{\boldsymbol{\alpha}\boldsymbol{\beta}}}\equiv 
\left(\begin{array}{cc}
\frac{\partial^{2} {S}_\xi}{\partial \alpha_{A}\partial \beta_{A}} &
\frac{\partial^{2} {S}_\xi}{\partial \alpha_{A}\partial \beta_{B}}\\
\frac{\partial^{2} {S}_\xi}{\partial \alpha_{B}\partial \beta_{A}}&
\frac{\partial^{2} {S}_\xi}{\partial \alpha_{B}\partial \beta_{B}}\\
\end{array}\right).
\nonumber
\end{equation}
In Eq.~(\ref{StoM}), we also define the auxiliary matrices $\mathbf{C}''\equiv {u''_{A}}^2\mathbf{A}''+ {u''_{B}}^2\mathbf{B}''$ and $\mathbf{D}''\equiv {v''_{A}}^2\mathbf{A}''+ {v''_{B}}^2\mathbf{B}''$, with
\begin{equation}
\mathbf{A}\equiv \frac{2 j}{(1+{u}_{A}{v}_{A})^2}\mathbf{I}_A
\quad\mathrm{and}\quad
\mathbf{B}\equiv \frac{2 j}{(1+{u}_{B}{v}_{B})^2}\mathbf{I}_B,
\nonumber
\end{equation}
where
\begin{equation}
\mathbf{I}_A\equiv\left(\begin{array}{ll}1&0\\0&0\end{array}\right)
\quad\mathrm{and}\quad
\mathbf{I}_B\equiv\left(\begin{array}{ll}0&0\\0&1\end{array}\right).
\nonumber
\end{equation}

To write the semiclassical formula of $\mathrm{K}_\xi$, we still need to define $\mathcal{G}_\xi=\mathcal{G}_\xi(\mathbf s_\eta^*,\mathbf s_\mu,T)$ and $\mathcal{D}_\xi=\mathcal{D}_\xi(\mathbf s_\eta^*,\mathbf s_\mu,T)$, such that
\begin{equation}
\begin{aligned}
\mathcal{G}_\xi =& \frac{i \hbar \xi}{4}\int_0^T 
\left[ \frac{\partial \dot u_A}{\partial u_A}-
\frac{\partial\dot v_A}{\partial v_A}+
\frac{\partial \dot u_B}{\partial u_B}-
\frac{\partial\dot v_B}{\partial v_B}
\right]\mbox{d}t,\\
\mathcal D_\xi =& \frac{\mbox{e}^{\tilde{\Lambda}}}{4j^2}~\det 
\left(\frac{i}{\hbar}
\mathbf{S}^{(\xi)}_{\mathbf{s}_\mu\mathbf{s}^*_\eta}
\right) .
\end{aligned}
\label{Gpref}
\end{equation}

Given all these functions, we finally write
\begin{equation}
\mathcal{K}_{\xi}\left( \mathbf s_{\eta}^*,\mathbf s_{\mu}, T \right)=
\sum_{\mathrm{c.t.}}\sqrt{\mathcal D_{\xi}}~
\mbox{e}^{\frac{i}{\hbar}\left(\mathcal{S_{\xi}}+\mathcal{G_{\xi}}\right)-\Lambda}  ,
\label{PP}
\end{equation}
where the term $\Lambda\equiv\Lambda_A+\Lambda_B$ is originated from the normalization of the states $|\mathbf{s}_\eta\rangle$ and $|\mathbf{s}_\mu\rangle$, with $\Lambda_A\equiv j\ln\left[(1+|s_{\eta A}|^2)(1+|s_{\mu A}|^2) \right]$ and the equivalent for $\Lambda_B$. The sum runs over all complex trajectories as defined earlier.

\subsection{Saddle point method}
\label{spm}

As already announced, we will now replace each quantum propagator seen in Eq.~(\ref{Pscs}) with its semiclassical version~(\ref{PP}). For clearness, we list below the replacements that we want to do,
\begin{equation}
\begin{aligned}
\langle s_A,s_B| \hat{U}_T|\mathbf{s}_{0}\rangle \to &~
\mathcal{K}_1(s_{A}^*,s_{B}^*;\mathbf{s}_{0};T), \\
\langle \mathbf{s}_{0}| \hat{U}_T^\dagger| \tilde{s}_{A},s_{B}\rangle \to &~
\mathcal{K}_2(\mathbf{s}_{0}^*; \tilde{s}_{A},s_{B};T), \\
\langle\tilde{s}_A,\tilde{s}_B| \hat{U}_T|\mathbf{s}_{0}\rangle \to &~
\mathcal{K}_3(\tilde{s}_A^*,\tilde{s}_B^*; \mathbf{s}_{0};T),\\          
\langle \mathbf{s}_{0}|\hat{U}_T^\dagger| s_{A},\tilde{s}_{B}\rangle \to  &~
\mathcal{K}_4(\mathbf{s}_{0}^*;s_{A},\tilde{s}_{B};T).
\end{aligned}
\label{KtoK}
\end{equation}
Notice that $\mathcal K_1$ and $\mathcal K_3$ refer to semiclassical forward propagators, while $\mathcal K_2$ and $\mathcal K_4$ are the backward ones. From this point on, we will refer to the trajectory involved in the calculation of $\mathcal K_k$ simply as $[\mathbf{u}_k(t),\mathbf{v}_k(t)]$, for $k=1,\ldots, 4$. In accordance with the (extended) complex phase space discussed in the last subsection, we will consider the pair $(s_A,s_A^*)$ as independent variables, and the same for $(\tilde{s}_A,\tilde{s}_A^*)$, $(s_B,s_B^*)$, and $(\tilde{s}_B,\tilde{s}_B^*)$. Thus, taking into account the boundary conditions~(\ref{bbnn}), we rewrite the integration variables of Eq.~(\ref{Pscs}) as
\begin{equation}
\begin{array}{llll}
 s_A^* \to v_{1A}'',& s_B^* \to v''_{1B},& 
 \tilde{s}_A \to u''_{2A},& s_B \to u''_{2B},\\ 
 \tilde{s}_A^* \to v''_{3A},& \tilde{s}_B^* \to v''_{3B},&
 s_A \to u''_{4A},&\tilde{s}_B \to u''_{4B}.
\end{array}
\end{equation}
Following this approach, it is also important to revisit Eq.~(\ref{NO1}), defined as an integration over the whole complex plane $s$. By considering $s$ and $s^*$ as independent variables, this integral is reinterpreted as two path integrals, one along the $s$-plane and the other along the $s^*$-plane. In this case, we have
\begin{equation}
\mbox{d}\mu(s) = \frac{2j+1}{2\pi i} \frac{\mbox{d}s~\mbox{d}s^*}{(1+s\,s^*)^2} ,
\nonumber
\end{equation}
a result that should be joined to our calculation. When we accomplish all these tasks, Eq.~(\ref{Pscs}) becomes
\begin{equation}
  P_T \approx  \int \sqrt{\mathcal{D}_1}\sqrt{\mathcal{D}_2}
  \sqrt{\mathcal{D}_3}\sqrt{\mathcal{D}_3}~\exp\big[\Phi\big]~\mbox{d}\boldsymbol{\mu},
\label{Psscs}
\end{equation}
where
\begin{equation}
\begin{aligned}
\Phi \equiv & \sum_{k=1}^4 
\left[\frac{i}{\hbar}\left(\mathcal{S}_k+\mathcal{G}_k\right)\right]-
\ln\left[(1+|s_{0A}|^2)(1+|s_{0B}|^2)\right]^{4j}\\
&-\ln\left(1+u''_{4A} v''_{1A}\right)^{2j}-\ln\left(1+u''_{2A} v''_{3A}\right)^{2j}\\
&-\ln\left(1+u''_{2B} v''_{1B}\right)^{2j}-\ln\left(1+u''_{4B} v''_{3B}\right)^{2j}
\end{aligned}
\nonumber
\end{equation}
and
\begin{equation}
\begin{aligned}
  \mbox{d}\boldsymbol{\mu}\equiv&
    \left(\frac{2j+1}{2\pi i}\right)^4
  \frac{\mbox{d}u''_{4A} \mbox{d}v''_{1A}}{(1+u''_{4A} v''_{1A})^2}
  \frac{\mbox{d}u''_{2A} \mbox{d}v''_{3A}}{(1+u''_{2A} v''_{3A})^2}\\
  &\times\frac{\mbox{d}u''_{2B} \mbox{d}v''_{1B}}{(1+u''_{2B} v''_{1B})^2}\,
  \frac{\mbox{d}u''_{4B} \mbox{d}v''_{3B}}{(1+u''_{4B} v''_{3B})^2}.
  \end{aligned}
\end{equation}
Functions $\mathcal D_k$, $\mathcal S_k$, and $\mathcal G_k$ clearly refer to their respective semiclassical propagator $\mathcal K_k$. Notice that the sum over complex trajectories was omitted in Eq.~(\ref{Psscs}), for simplicity. This point will be resumed later.

We are now ready to deduce a semiclassical expression for the linear entropy, attacking Eq.~(\ref{Psscs}) through the saddle point method (or steepest descent method)~\cite{bleistein}. To start the computation, we first recognize that the line integral~(\ref{Psscs}) is defined in a space of eight complex variables, which we will rewrite as 
\begin{equation}
{\mathbf{r}}^{\mathrm{T}}\equiv
( v_{1A}'',v''_{1B}, u''_{2A},   u''_{2B}, v''_{3A}, v''_{3B},  u''_{4A}, u''_{4B}),
\end{equation}
where $\mathbf{r}^{\mathrm{T}}$ indicates the transpose of the column vector~$\mathbf{r}$. For each point of the integration path, the input parameters of all~$\mathcal{K}_k$ are automatically determined, which define the four trajectories needed to evaluate the integrand. Then, the direct prescription to compute integral~(\ref{Psscs}) is following the path of integration and summing the contribution of each point. 

However, it happens that the integrand of Eq.~(\ref{Psscs}) fastly oscillates around zero along any generic path. It occurs because of the semiclassical regime assumed here: as $j\to\infty$ (with $\hbar\sim 1/j$), a simple inspection in the function $\mathcal{S}_k$, present in $\Phi$, assures this behavior. Therefore, one can say that the integral vanishes along generic paths, which is actually the reason why we have so far neglected any information about paths of integration. The purpose of the steepest descent method consists of finding the saddle points of the integrand and, supported by Cauchy's integral theorem, performing the integral along its steepest descents. By doing so, the rapid oscillations are dropped out because the imaginary part of $\Phi$ is constant along this particular path~\cite{bleistein}. Moreover, in the regime considered, in general, it is enough to replace $\Phi$ with its second order expansion around the saddle point, so that solving Eq.~(\ref{Psscs}) simply becomes computing a Gaussian integral.

The saddle point $\bar{\mathbf{r}}$ is given by the solution of $\nabla\Phi=0$, where the derivatives are taken with respect to the components of $\mathbf{r}$. In the semiclassical limit, the derivatives of $\mathcal{G}_k$ can be disregarded in comparison to other terms of~$\Phi$. Therefore, with the help of Eq.~(\ref{variaS}), we find that the saddle point should satisfy
\begin{equation}
\begin{array}{llll}
\bar{v}''_{1A}=\bar{v}''_{4A}, &  \bar{v}''_{1B}=\bar{v}''_{2B}, &
\bar{u}''_{2A}=\bar{u}''_{3A}, &  \bar{u}''_{2B}=\bar{u}''_{1B},\\
\bar{v}''_{3A}=\bar{v}''_{2A}, &  \bar{v}''_{3B}=\bar{v}''_{4B}, &
\bar{u}''_{4A}=\bar{u}''_{1A}, &  \bar{u}''_{4B}=\bar{u}''_{3B}.
\end{array}
\label{fbc}
\end{equation}
Notice that these relations imply that the final points of the four critical trajectories are mutually connected. For instance, the final point $(\bar{u}''_{1A},\bar{u}''_{1B},\bar{v}''_{1A},\bar{v}''_{1B})$ of the trajectory entering in $\mathcal{K}_1$ must be equal to $(\bar{u}''_{4A},\bar{u}''_{2B},\bar{v}''_{4A},\bar{v}''_{2B})$, which represents a joint constraint with the final points of the trajectories 2 and 4 (analogous relations can be found for other trajectories). Because of this property, we also call these four trajectories, used to evaluate the semiclassical linear entropy, as {\em entangled-boundary-condition trajectories}, in accordance with the nomenclature adopted in Ref.~\cite{arlans}.

The eight equalities of Eq.~(\ref{fbc}), in addition to the initial conditions
\begin{equation}
\begin{array}{l}
\bar{\mathbf{u}}_1'=\bar{\mathbf{u}}_3'=\mathbf{s}_0
\quad\mathrm{and}\quad
\bar{\mathbf{v}}_2'=\bar{\mathbf{v}}_4'=\mathbf{s}_0^*, 
\end{array}
\label{inibc}
\end{equation}
give all prescriptions needed to find contributing sets of four complex trajectories. In particular, when $\bar{\mathbf{v}}_1'=\bar{\mathbf{v}}_3'=\mathbf{s}_0^*$ and $\bar{\mathbf{u}}_2'=\bar{\mathbf{u}}_4'=\mathbf{s}_0$, all trajectories are the same and real, as dicussed in Sect.~\ref{sp}. As studied in Refs.~\cite{pra2010,pra2012}, these trajectories also satisfy Eq.~(\ref{fbc}), leading to a semiclassical approximation for the linear entropy which agrees with the quantum result only for short evolution time. As previously mentioned, here we will get a better accuracy by including the complex trajectories. 

Once the saddle point of the integrand is understood, we proceed with the calculation, performing the expansion of the integrand around it. As usual, functions $\mathcal{G}_k$ and prefactors $\mathcal{D}_k$ are just calculated at the saddle point, so that
\begin{equation}
P_T \approx \left( \frac{2j+1}{2\pi i}\right)^4  
\left(\prod_{k=1}^4
\frac{\sqrt{{\bar{\mathcal{D}}_k}}}{\bar{\mathcal{J}_{k}}}\right)
\mbox{e}^{\bar\Phi}I_{\mathrm{G}},
\label{pureza3}
\end{equation}
where $\bar{\mathcal{J}}_k = (1+\bar{u}''_{kA}\bar{v}''_{kA})(1+\bar{u}''_{kB}\bar{v}''_{kB})$ and $I_{\mathrm{G}}$ is the Gaussian integral
\begin{equation}
I_{\mathrm{G}}\equiv \int \mbox{d}^8 \mathbf{r}~
\exp\left[-\frac12\mathbf{r}^{\mathrm T}~\bar{\mathbf{Q}}~\mathbf{r}\right]
=\sqrt{\frac{(2\pi)^8}{\det{\bar{\mathbf{Q}}}}}.
\nonumber
\end{equation}
The bar over the functions indicates that they should be evaluated with the saddle point $\bar{\mathbf{r}}$, and $\bar{\mathbf{Q}}$ is the $8\times8$ matrix
\begin{equation}
\bar{\mathbf{Q}}\equiv-\left[
\begin{array}{cccc}
\bar{\mathbf{R}}^{(1)}_{\mathbf{vv}}  &  -\bar{\mathbf{B}}^{(2)}  & \mathbf{0} & -\bar{\mathbf{A}}^{(4)} \\
-\bar{\mathbf{B}}^{(1)}  & \bar{\mathbf{R}}^{(2)}_{\mathbf{uu}}  &  -\bar{\mathbf{A}}^{(3)}  & \mathbf{0} \\
\mathbf{0}  & -\bar{\mathbf{A}}^{(2)}  &  \bar{\mathbf{R}}^{(3)}_{\mathbf{vv}}  & -\bar{\mathbf{B}}^{(4)}  \\
-\bar{\mathbf{A}}^{(1)}  & \mathbf{0}  & -\bar{\mathbf{B}}^{(3)}  &  \bar{\mathbf{R}}^{(4)}_{\mathbf{uu}} \\
\end{array}\right],
\label{matriz_q}
\end{equation}
where
\begin{equation}
\bar{\mathbf{R}}^{(k)}_{\mathbf{vv}}\equiv
\bar{\mathbf{C}}^{(k)}+
\frac{i}{\hbar}\bar{\mathbf{S}}^{(k)}_{\mathbf{v}''\mathbf{v}''}
\quad\mathrm{and}\quad
\bar{\mathbf{R}}^{(k)}_{\mathbf{uu}}\equiv
\bar{\mathbf{D}}^{(k)}+
\frac{i}{\hbar}\bar{\mathbf{S}}^{(k)}_{\mathbf{u}''\mathbf{u}''}.
\nonumber
\end{equation}
The definition of $\bar{\mathbf{A}}^{(k)}$, $\bar{\mathbf{B}}^{(k)}$, $\bar{\mathbf{C}}^{(k)}$, $\bar{\mathbf{D}}^{(k)}$, and $\bar{\mathbf{S}}^{(k)}_{\boldsymbol{\alpha\beta}}$ are presented right below Eq.~(\ref{StoM}), provided with information about the trajectory number~$k$.

Equation~(\ref{pureza3}) can be substantially simplified if we replace the second derivatives of the complex action with the elements of the stability matrix $\mathbf{M}$, according to Eq.~(\ref{StoM}). Using this strategy, we finally get the semiclassical linear entropy of the reduced state derived from the pure state $|\psi_0\rangle$, given by Eq.~(\ref{s0}), as a function of the time evolution $T$
\begin{equation}
 S_{\mathrm{sc}} (T) = 1-
  \sum_{\mathrm{sets}} \sqrt{\frac{\mathcal{A}}{\det\mathbf{F}}}~ 
  \mbox{e}^{\frac{i}{\hbar}\left[F_1-F_2+F_3-F_4\right]}.
  \label{Slinsemi}
\end{equation}
Here, for simplicity, we remove the bar over the symbols and define the matrix
\begin{equation}
\mathbf{F} \equiv
\left(\begin{array}{cccc}
-\mathbf{M}_{\mathbf{uv}}^{(1)}&\mathbf{I}_B\mathbf{M}_{\mathbf{uu}}^{(2)}&
0&\mathbf{I}_A\mathbf{M}_{\mathbf{uu}}^{(4)}\\
\mathbf{I}_B\mathbf{M}_{\mathbf{vv}}^{(1)}&-\mathbf{M}_{\mathbf{vu}}^{(2)}&
\mathbf{I}_A\mathbf{M}_{\mathbf{vv}}^{(3)}&0\\
0&\mathbf{I}_A\mathbf{M}_{\mathbf{uu}}^{(2)}&
-\mathbf{M}_{\mathbf{uv}}^{(3)}&\mathbf{I}_B~\mathbf{M}_{\mathbf{uu}}^{(4)}\\
\mathbf{I}_A\mathbf{M}_{\mathbf{vv}}^{(1)}&0&
\mathbf{I}_B\mathbf{M}_{\mathbf{vv}}^{(3)}&-\mathbf{M}_{\mathbf{vu}}^{(4)}
\end{array}\right),
\nonumber
\end{equation}
and the functions  
\begin{equation}
F_k\equiv \int^T_0
          \left( i\hbar j \chi_k - \tilde{H} \right)\mbox{d}t 
          +\mathcal{G}_k 
\nonumber
\end{equation}
and $\mathcal{A}\equiv\mathcal{A}_A\mathcal{A}_B$, with
\begin{equation}
\mathcal{A}_A\equiv\prod_{k=1}^4 
                    \frac{1+u''_{kA}v''_{kA}}{1+u'_{kA}v'_{kA}}
                    \left(\frac{1+u'_{kA}v'_{kA}}{1+|s_{0A}|^2}\right)^{2j},
\nonumber
\end{equation}
and the equivalent for $\mathcal{A}_B$. 

Equation~(\ref{Slinsemi}) is the main result of the present paper and an example of its application will be presented in Sect.~\ref{example1}. The sum in $ S_{\mathrm{sc}} (T) $ indicates that all sets of classical trajectories respecting the boundary conditions~(\ref{fbc}) and~(\ref{inibc}) are, in principle, important to approach the quantum linear entropy~(\ref{Sdef}). Notice that this consideration recovers the arbitrary exclusion of the summation in Eq.~(\ref{Psscs}). However, we need to comment that numerical evidence shows that some sets of trajectories furnish unphysical results which give origin to unexpected divergent behaviors, for example. These contributions will be arbitrarily excluded from the calculation. Although we do not mathematically prove this argument, we associate this issue with those saddle points of Eq.~(\ref{Psscs}) whose steepest descents cannot be deformed from the original path of integration. We remind that this kind of problem is very common in applications of coherent state propagators~\cite{ribeiro04, ribeiro05}. 
 
The derivation of $S_{\mathrm{sc}} (T)$ assumes the semiclassical regime, as we have considered large values of $j$. Working with the extremal case $j\to\infty$ and based on the correspondence principle, we can say that real trajectories are enough to reproduce the quantum behavior. By relaxing this condition, we expect that complex trajectories become important to the approximation. Then, we can think that the complex contributions closer to the real one should be among the most important to evaluate $S_{\mathrm{sc}} (T)$. For this reason, in the next section, we investigate general conditions needed for finding sets of quasi-real trajectories, as an attempt to understanding their origin.

\section{Sets of quasi-real trajectories}
\label{ntv}

A natural question that arises from the present theory concerns the investigation of the physical mechanism behind the emergence of complex classical trajectories contributing to Eq.~(\ref{Slinsemi}). The comprehension of this process will clarify aspects of the quantum-classical connection related to the entanglement phenomenon. As an effort to unravel this issue, in the present section, we will explore the vicinity of the real contributing trajectory, searching for complex trajectories satisfying the boundary conditions~(\ref{fbc}) and~(\ref{inibc}). 

To follow this idea, we remind that, for any input parameters $\mathbf{s}_0$ and $T$, we know that the set of four real and identical trajectories starting from $\bar{\mathbf{u}}'=\mathbf{s}_0$ and $\bar{\mathbf{v}}' = \mathbf{s}_0^*$ contributes to Eq.~(\ref{Slinsemi}). In this section, a bar over the symbol refers to the real trajectory. We will look for four complex trajectories $\mathbf{w}_k(t)\equiv[\mathbf{u}_k(t),\mathbf{v}_k(t)]$, for $k=1,\ldots,4$, close to the real one and constrained to Eqs.~(\ref{fbc}) and (\ref{inibc}), with the former rewritten here as
\begin{equation}
\begin{array}{lll}
\mathbf{w}''_1 &=& 
\left(\begin{array}{cc}\mathbf{I}_A&0\\0&\mathbf{I}_A\end{array}\right)\mathbf{w}''_4+
\left(\begin{array}{cc}\mathbf{I}_B&0\\0&\mathbf{I}_B\end{array}\right)\mathbf{w}''_2,\\\\
\mathbf{w}''_3 &=& 
\left(\begin{array}{cc}\mathbf{I}_A&0\\0&\mathbf{I}_A\end{array}\right)\mathbf{w}''_2+
\left(\begin{array}{cc}\mathbf{I}_B&0\\0&\mathbf{I}_B\end{array}\right)\mathbf{w}''_4.
\end{array}
\label{fbcc}
\end{equation}
As all these trajectories are in the vicinity of the real one, we have 
\begin{equation}
\mathbf{w}_k(t) = \bar{\mathbf{w}}(t)+\delta\mathbf{w}_k(t),
\quad\mathrm{for}\quad k=1,\ldots,4 .
\end{equation}
where $\delta\mathbf{w}_k(t)$, by construction, are small complex numbers. Notice that, if we were able to calculate all $\delta\mathbf{w}_k'$ through the imposition of the initial and final boundary conditions, we get a new set of four contributing trajectories. Imposing the initial constraints implies 
\begin{equation}
\delta u_{1A}'=\delta u_{3A}'=\delta v_{2A}'=\delta v_{4A}'=0,
\end{equation}
and the same for part $B$. Notice that $\delta v_{1A}'$, $\delta v_{1B}'$, $\delta v_{3A}'$, $\delta v_{3B}'$, $\delta u_{2A}'$, $\delta u_{2B}'$, $\delta u_{4A}'$, and $\delta u_{4B}'$ are still undetermined. To solve these variables we need to work with conditions~(\ref{fbcc}). Summing and subtracting them, we get, respectively,
\begin{equation}
\begin{aligned}
\delta\mathbf{w}_{1+3}'' &= \delta\mathbf{w}_{2+4}'', \\
\delta\mathbf{w}_{1-3}'' &=
\left(\begin{array}{cc}
\mathbf{I}_B-\mathbf{I}_A&0\\0&\mathbf{I}_B-\mathbf{I}_A 
\end{array}\right) \delta\mathbf{w}_{2-4}'',
\end{aligned}
\end{equation}
where $\delta\mathbf{w}_{1 \pm 3}\equiv\delta\mathbf{w}_1 \pm \delta\mathbf{w}_3$ and $\delta\mathbf{w}_{2 \pm 4}\equiv\delta\mathbf{w}_2 \pm \delta\mathbf{w}_4$.

The next step is to write the final displacements as functions of the initial ones using the stability matrix~(\ref{stabmat}), finding
\begin{equation}
\begin{aligned}
\mathbf{\bar{M}_{uv}}\delta\mathbf{v}_{1+3}' &= \mathbf{\bar{M}_{uu}}\delta\mathbf{u}_{2+4}', \\
\mathbf{\bar{M}_{vv}}\delta\mathbf{v}_{1+3}' &= \mathbf{\bar{M}_{vu}}\delta\mathbf{u}_{2+4}', \\
\mathbf{\bar{M}_{uv}}\delta\mathbf{v}_{1-3}' &=
\left(\begin{array}{cccc}-1&0\\0&1\end{array}\right)\mathbf{\bar{M}_{uu}}\delta\mathbf{u}_{2-4}', \\
\mathbf{\bar{M}_{vv}}\delta\mathbf{v}_{1-3}' &=
\left(\begin{array}{cccc}-1&0\\0&1\end{array}\right)\mathbf{\bar{M}_{vu}}\delta\mathbf{u}_{2-4}'.
\end{aligned}
\label{eq32}
\end{equation}
By manipulating the first two equations, we get
\begin{equation}
\begin{aligned}
&\left[ \mathbf{\bar{M}_{vv}} -\mathbf{\bar{M}_{vu}}\mathbf{\bar M}_{\mathbf{uu}}^{-1}\mathbf{\bar{M}_{uv}}\right]
\delta\mathbf{v}_{1+3}' \equiv \mathbf{\bar{M}}_{13}^+ \delta\mathbf{v}_{1+3}'= 0,\\
&\left[ \mathbf{\bar{M}_{uu}} -\mathbf{\bar{M}_{uv}}\mathbf{\bar M}_{\mathbf{vv}}^{-1}\mathbf{\bar{M}_{vu}}\right]
\delta\mathbf{u}_{2+4}' \equiv \mathbf{\bar{M}}_{24}^+ \delta\mathbf{v}_{2+4}'= 0,
\end{aligned}
\label{plus}
\end{equation}
where the former (latter) assumes that $\mathbf{\bar M}_{\mathbf{uu}}$ ($\mathbf{\bar M}_{\mathbf{vv}}$) is invertible. The other two equations of~(\ref{eq32}) furnish
\begin{equation}
\begin{aligned}
&\left[\mathbf{\bar{M}_{vv}}-
\mathbf{\bar{M}}_{\mathbf{vu}}^{\star}
\mathbf{\bar M}_{\mathbf{uu}}^{-1}
\mathbf{\bar{M}}_{\mathbf{uv}}^{\star}
\right]
\delta \mathbf{v}_{1-3}' \equiv \mathbf{\bar{M}}_{13}^- \delta\mathbf{v}_{1-3}' =0,\\
&\left[\mathbf{\bar{M}_{uu}}-
\mathbf{\bar{M}}_{\mathbf{uv}}^{\star}
\mathbf{\bar M}_{\mathbf{vv}}^{-1} 
\mathbf{\bar{M}}_{\mathbf{vu}}^{\star}
\right]\delta \mathbf{u}_{2-4}' \equiv \mathbf{\bar{M}}_{24}^- 
\delta\mathbf{v}_{2-4}'= 0,
\end{aligned}
\label{minus}
\end{equation}
where we define
\begin{equation}
\mathbf{\bar{M}}_{\mathbf{uv}}^{\star}\equiv
\left(\begin{array}{cc}-1&0\\0&1\end{array}\right)
\mathbf{\bar{M}}_{\mathbf{uv}},
\nonumber
\end{equation}
and the equivalent for $\mathbf{\bar{M}}_{\mathbf{vu}}^{\star}$. Using the identities
\begin{equation}
\det\mathbf{\bar{M}}=\left\{
\begin{array}{lll}
\det\left[\mathbf{\bar{M}_{uu}}\mathbf{\bar{M}_{13}^+}\right],&& \mathrm{for}\; \det\mathbf{\bar{M}_{uu}}\neq 0,\\
\det\left[\mathbf{\bar{M}_{vv}}\mathbf{\bar{M}_{24}^+}\right],&& \mathrm{for}\; \det\mathbf{\bar{M}_{vv}}\neq 0,
\end{array}\right.
\label{schur}
\end{equation}
according to Eq.~(\ref{plus}), we conclude that non-trivial solutions of $\delta\mathbf{v}_{1+3}'$ and $\delta\mathbf{v}_{2+4}'$ will exist only if $\det\mathbf{\bar{M}}=0$. However, it can be shown (see Ref.~\cite{pra2012}, appendix C) that 
\begin{equation}
\det\mathbf{\bar{M}}=
\frac{\left(1+ \bar{u}''_{A} \bar{v}''_{A} \right)^2\left(1+ \bar{u}''_{B} \bar{v}''_{B} \right)^2}
{\left(1+ \bar{u}'_{A} \bar{v}'_{A} \right)^2\left(1+ \bar{u}'_{B} \bar{v}'_{B} \right)^2} \neq 0,
\end{equation}
as $\mathbf{\bar{M}}$ is the stability matrix of a real trajectory, that is, $\bar{\mathbf{u}}=\bar{\mathbf{v}}^*$. Therefore, the solution of Eq.(\ref{eq32}) is 
\begin{equation}
\delta\mathbf{v}_{1+3}'=\delta\mathbf{v}_{2+4}'=0
\Longrightarrow  \left\{
\begin{array}{lll}
\delta\mathbf{v}_{1}'&=&-\delta\mathbf{v}_{3}',\\
\delta\mathbf{v}_{2}'&=&-\delta\mathbf{v}_{4}'.
\end{array}\right.
\label{solplus}
\end{equation}
Concerning the other two variables $\delta\mathbf{v}_{1-3}'$ and $\delta\mathbf{v}_{2-4}'$, in analogy to Eq.~(\ref{schur}), we notice that
\begin{equation}
\det\mathbf{\bar{M}}^{\star}=\left\{
\begin{array}{lll}
\det\left[\mathbf{\bar{M}_{uu}}\mathbf{\bar{M}_{13}^-}\right],&& \mathrm{for}\; \det\mathbf{\bar{M}_{uu}}\neq 0,\\
\det\left[\mathbf{\bar{M}_{vv}}\mathbf{\bar{M}_{24}^-}\right],&& \mathrm{for}\; \det\mathbf{\bar{M}_{vv}}\neq 0,
\end{array}\right.
\end{equation}
where we define
\begin{equation}
\mathbf{\bar{M}}^{\star}\equiv
\left(\begin{array}{cc}
\mathbf{\bar{M}_{uu}} & \mathbf{\bar{M}_{uv}}^\star\\
\mathbf{\bar{M}_{vu}}^\star & \mathbf{\bar{M}_{vv}}
\end{array}\right).
\label{Mstar}
\end{equation}
Therefore, according to Eq.~(\ref{minus}), the condition for finding non-trivial solutions of $\delta\mathbf{v}_{1-3}'$ and $\delta\mathbf{v}_{2-4}'$ is $\det \mathbf{\bar{M}}^{\star} = 0$. Contrarily to $\bar{\mathbf{M}}$, for matrix~(\ref{Mstar}) we  
have not found a general result for its determinant, so that we should analyze this point for each application. For the {\em particular case} of a non-singular $\mathbf{\bar{M}}^{\star}$, we have  
\begin{equation}
\delta\mathbf{v}_{1-3}'=\delta\mathbf{v}_{2-4}'=0
\Longrightarrow  \left\{
\begin{array}{lll}
\delta\mathbf{v}_{1}'&=&\delta\mathbf{v}_{3}',\\
\delta\mathbf{v}_{2}'&=&\delta\mathbf{v}_{4}'.
\end{array}\right.
\end{equation}
Then, according to Eq.~(\ref{solplus}), we find $\delta \mathbf{w}_k'=0$ for all $k$, implying the absence of complex contributing trajectories arbitrarily close to the real one. This will be the case of the system studied in the next section.

At last, we highlight that our efforts to probe the vicinity of a real contributing trajectory are justified because it is expected to be the most important phase-space region, provided that the semiclassical regime is assumed. Although we still have no clear understanding of how complex contributing trajectories could be continuously originated from the real phase space, we found the condition for it occurs, which is $\det\bar{\mathbf{M}}^\star=0$.

\begin{figure}[!t]
\centerline{
\includegraphics[width=8cm,angle=0]{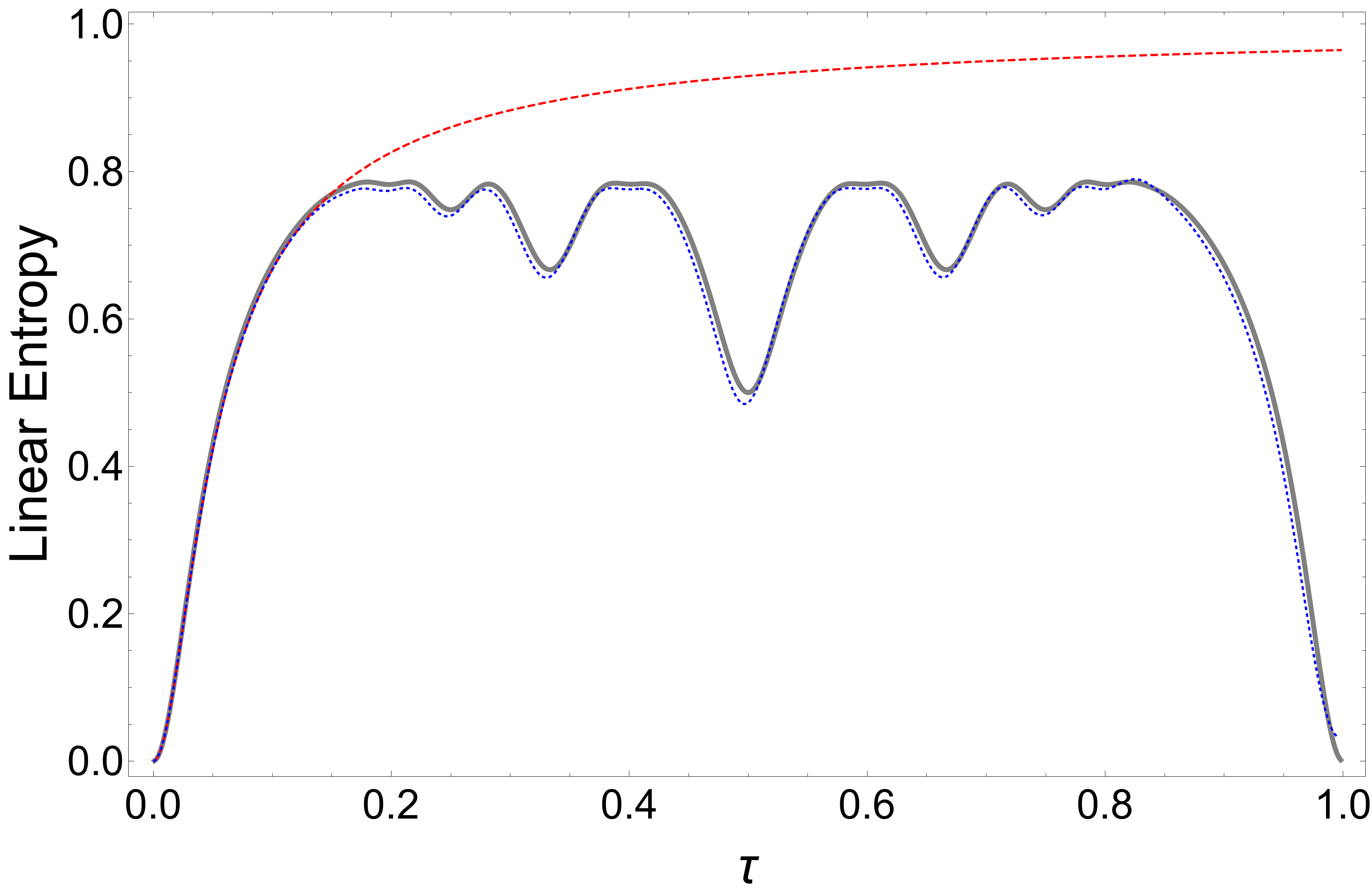}}
\caption{The gray solid line illustrates the quantum linear entropy~(\ref{SH2spins}) as a function of the dimensionless time $\tau$. We also show results for the semiclassical entropy~(\ref{Slinsemi}): the red dashed line illustrates $S_{\mathrm{sc}}$  when we take into account only real trajectories, while the blue dotted line contemplates the inclusion of a few dozens of complex sets. The numerical parameters used to build this figure are given by Eq.~(\ref{nv}).}
\label{f1}
\end{figure}

\section{Phase coupling Hamiltonian}
\label{example1}

\begin{figure*}[!t]
\centerline{
\includegraphics[width=5cm,angle=0]{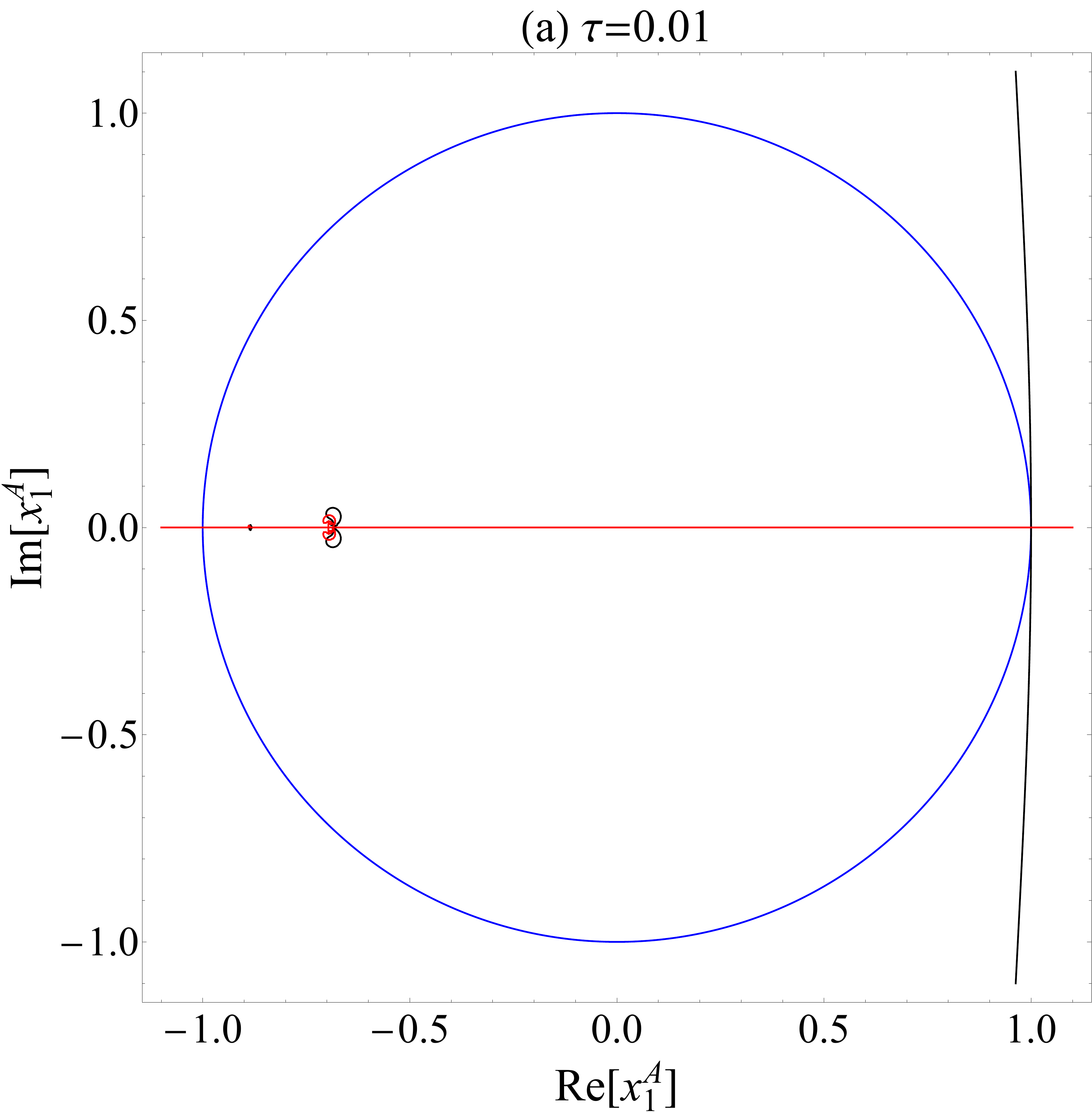}
\includegraphics[width=5cm,angle=0]{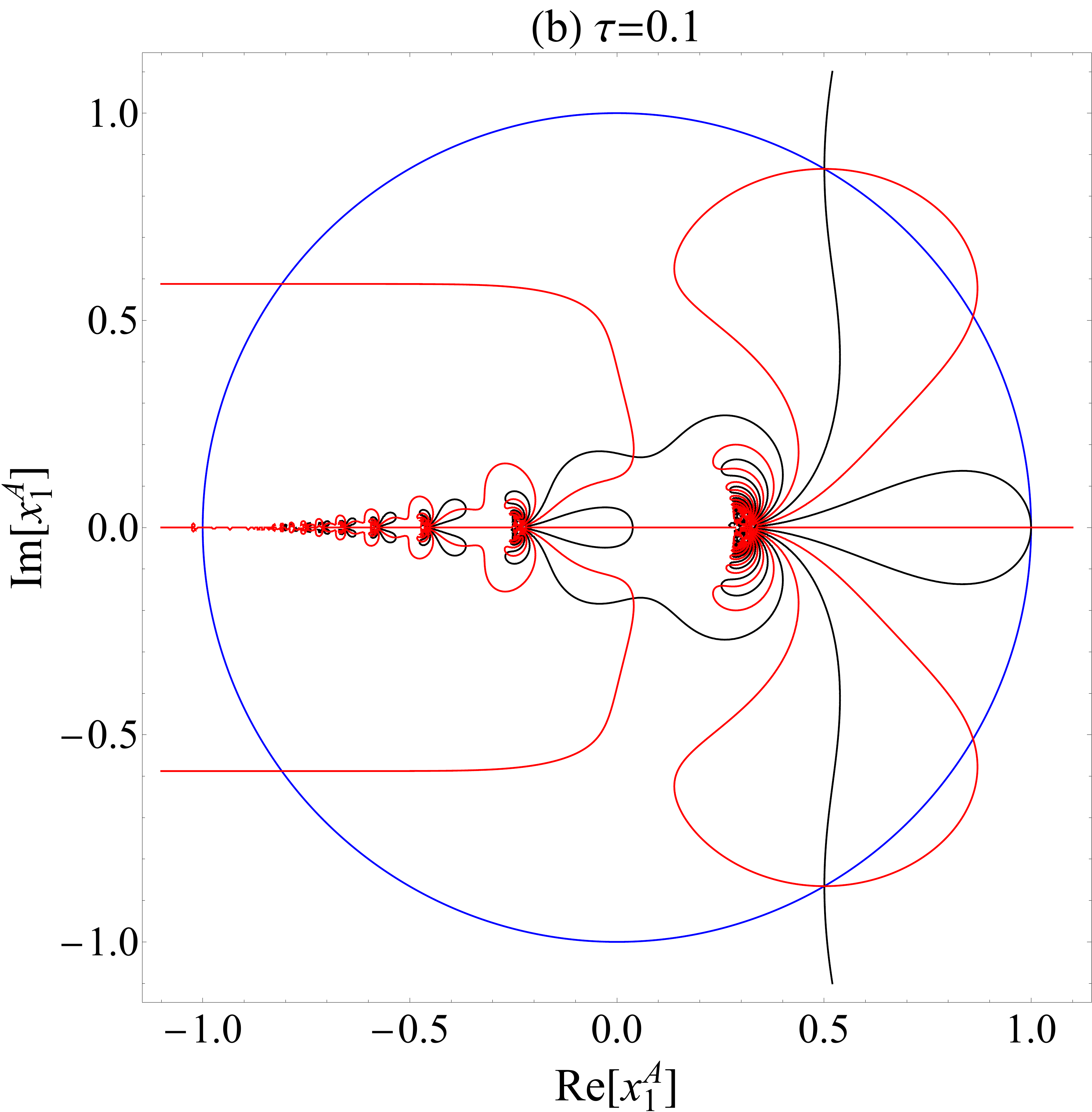}
\includegraphics[width=5cm,angle=0]{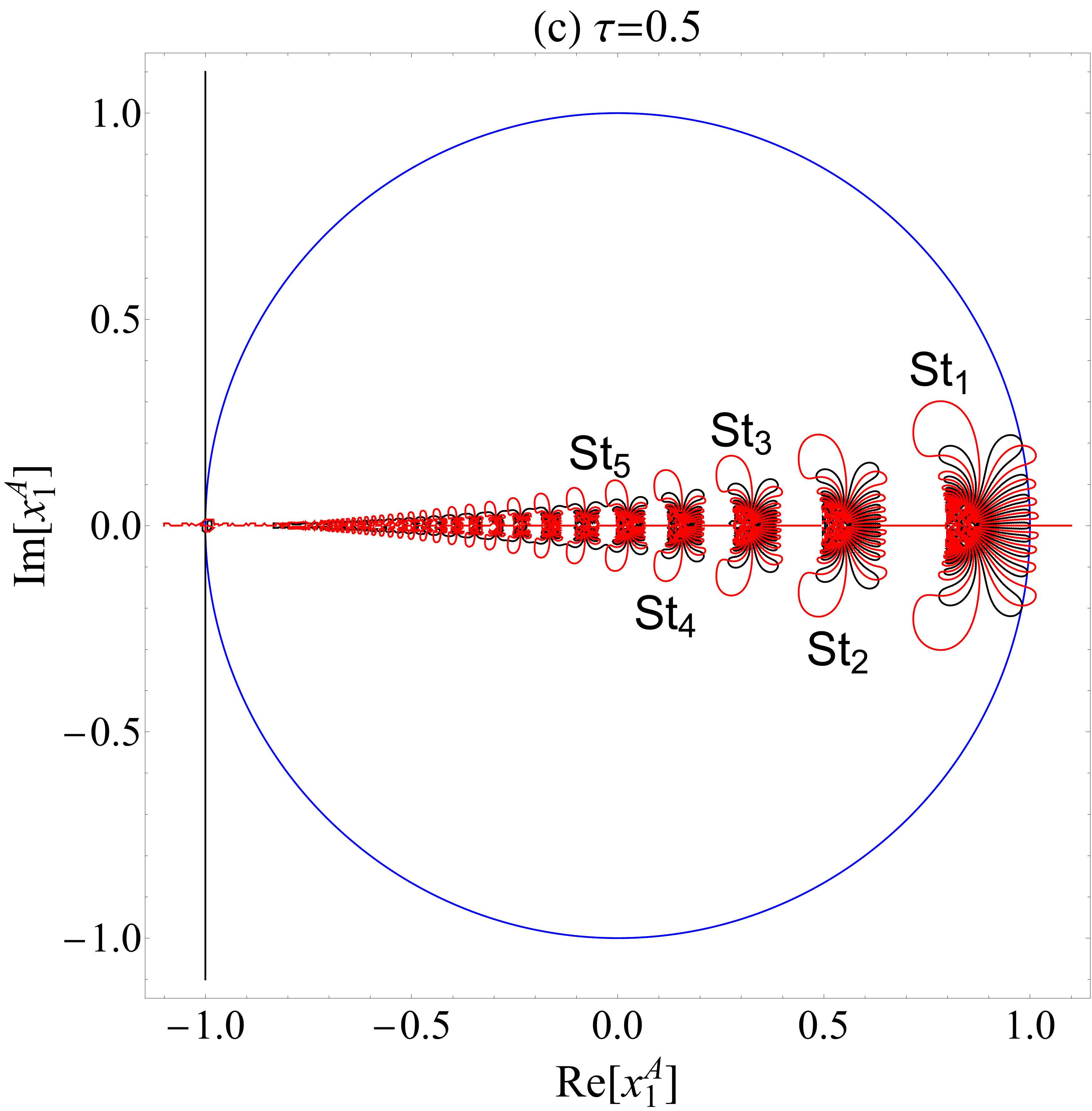}}
\caption{Contour plots of $f(x_1^A)$ in the $x_1^A$-complex plane. Black lines refer to the curve where $\mathrm{Re}[f(x_1^A)]=0$, while the red ones refer to $\mathrm{Im}[f(x_1^A)]=0$. The blue curve represents the unitary circle. Solutions of Eq.~(\ref{eqtrans}) are given by the intersections of black and red lines. In panels (a)-(c), the value of $\tau$ was chosen to be 0.01, 0.1, and 0.5, respectively. The other numerical parameters are shown in Eq.~(\ref{nv}). A magnification of structures $\mathrm{St}_1$,\ldots, $\mathrm{St}_4$, seen in panel (c), is shown in Fig.~\ref{f3}.}
\label{f2}
\end{figure*}

Our first application of the present semiclassical theory concerns a system of two particles, $A$ and $B$, whose spins interact with each other according to the Hamiltonian
\begin{equation} 
\hat{H}_{\mathrm{pc}}=\lambda \hbar [\hat{J}_A^{{(z)}} \otimes \hat{J}_B^{{(z)}}],
\label{H2spins}
\end{equation} 
where $\hat{\mathbf{J}}_A=(\hat{J}_A^{{(x)}}, \hat{J}_A^{{(y)}}, \hat{J}_A^{{(z)}})$ is the spin operator acting on $A$ (the same for part $B$) and $\lambda$ is the coupling constant. This example was already used in Ref.~\cite{pra2012}, but limited to the case where only real trajectories were used to compute the semiclassical linear entropy~(\ref{Slinsemi}). 

By considering the initial state~(\ref{s0}), the quantum entropy~(\ref{Sdef}) for this system becomes
\begin{equation}
S_{\mathrm{pc}}=1- \sum c^{(2j)}_{n_A} c^{(2j)}_{m_A} c^{(2j)}_{n_B} c^{(2j)}_{m_B}
\mbox{e}^{-i\lambda T\delta_A \delta_B},
\label{SH2spins}
\end{equation}
where $\delta_A \equiv n_A-m_A$ and
\begin{equation}
c^{(2j)}_{n_A} \equiv \binom{2j}{n_A}~\frac{|s_{0A}|^{\sigma_A}}{(1+|s_{0A}|^2)^{2j}},
\nonumber
\end{equation}
with $\sigma_A \equiv n_A-m_A$, and the analog for part $B$. The sum is over $n_A$, $m_A$, $n_B$, and $m_B$, running from 0 to $2j$. As Eq.~(\ref{SH2spins}) is clearly periodic in $T$, with a period $T_\mathrm{r}\equiv 2\pi/\lambda$, it is convenient to define the dimensionless time 
\begin{equation}
\tau\equiv {T}/{T_{\mathrm{r}}},
\nonumber
\end{equation}
and restrict our study to the interval $0<\tau<1$. In Fig.~\ref{f1}, for numerical values chosen as 
\begin{equation}
j=4.5\quad\mathrm{and}\quad s_{0A}=s_{0B}=\lambda=1,
\label{nv}
\end{equation}
we illustrate (black solid line) the behavior of the quantum entanglement dynamics~(\ref{SH2spins}) during a period $T_{\mathrm{r}}$. As expected, for the initial separable state, $S_{\mathrm{pc}}$ is null, growing up as time increases. Then, after some oscillations for intermediate values of time, it returns to zero for $\tau=1$.

In order to obtain the semiclassical linear entropy to compare with the quantum calculation, we need to consider the equivalent classical description, whose Hamiltonian function~(\ref{Htilde}) is given by
\begin{equation}
\tilde{H}_{\mathrm{pc}}(\mathbf{u}, \mathbf{v})=
\lambda\hbar j^2
\left( \frac{1-u_{A} v_{A}}{1+u_{A} v_{A} } \right)
\left( \frac{1-u_{B} v_{B}}{1+u_{B} v_{B} } \right).
\nonumber
\end{equation}
Therefore, the equations of motion (\ref{emnn}) can be easily solved, so that the trajectories, written in terms of generic initial conditions $\mathbf{u}'$ and $\mathbf{v}'$, are given by 
\begin{equation}
\begin{array}{ll}
u_{A}(t)=u_{A}'\,\mbox{e}^{+ \lambda_Bt}, &
\quad u_{B}(t)=u_{B}'\,\mbox{e}^{+ \lambda_At},\\
v_{A}(t)=v_{A}'\,\mbox{e}^{- \lambda_Bt}, &
\quad v_{B}(t)=v_{B}'\,\mbox{e}^{- \lambda_At},
\end{array}
\label{traj1}
\end{equation}
where 
\begin{equation}
\lambda_A\equiv i\lambda j \left( \frac{1-u'_{A} v'_{A}}{1+u'_{A} v'_{A}}\right),
\nonumber
\end{equation} 
and the same for $\lambda_B$. We point out that, if one differentiates Eq.~(\ref{traj1}), the stability matrix~(\ref{stabmat}), which is an important ingredient of Eq.~(\ref{Slinsemi}), can be easily achieved~\cite{pra2012}.

After this brief presentation of the classical description of the problem, we can finally look for the sets of four entangled trajectories $[\mathbf{u}_k(t),\mathbf{v}_k(t)]$, with $k=1,\ldots,4$, which contribute to $S_{\mathrm{sc}}$. Notice that the initial boundary conditions~(\ref{inibc}) are easily imposed to the trajectories~(\ref{traj1}). The application of the eight final constraints~(\ref{fbc}), on the other hand, requires extensive but straightforward algebra. The strategy to deal with this point consists of writing the unknown initial variables as
\begin{equation}
\begin{array}{ll}
v_{1A}'= x_1^A s_{0A}^*, &\quad u_{2A}'= x_2^A s_{0A},\\
v_{3A}'= x_3^A s_{0A}^*, &\quad u_{4A}'= x_4^A s_{0A},
\end{array}
\label{nvx}
\end{equation}
and the equivalent for $B$. With these expressions, by imposing Eq.~(\ref{fbc}), we get a system of eight variables ($x_1^A,\ldots,x_4^A,x_1^B,\ldots,x_4^B$), and the same number of equations. By manipulating them, one can show that the variable $x_1^A$ should be a solution of the transcendental equation
\begin{equation}
f(x_1^A) \equiv f_B\left[ f_A (x_1^A)\right]-x_1^A=0,
\label{eqtrans}
\end{equation}
where $f_A$ (analogously for $f_B$) is a function defined according to
\begin{equation}
 f_A(x)=\exp{\left[ 
 \frac{-2ij\lambda|s_{0A}|^2 T (x^2-1)}
 {(1+|s_{0A}|^2 x)(x+|s_{0A}|^2 )} \right]}.
\nonumber
\end{equation}
Once Eq.~(\ref{eqtrans}) is numerically solved for $x_1^A$, the variable $x_1^B$ can be obtained from $x_1^B=f_A(x_1^A)$, and the other six variables are given by
\begin{equation}
\begin{array}{lll}
x_2^A=1/x_1^A, &\quad x_3^A=1/x_1^A, &\quad x_4^A=x_1^A,\\
x_2^B=x_1^B, &\quad x_3^B=1/x_1^B, &\quad x_4^B=1/x_1^B.
\end{array}
\nonumber
\end{equation}  
With the values of all these variables in hand, we return to Eq.~(\ref{nvx}) to get the rest of the information needed to find the initial points, $\mathbf{u}_k'$ and $\mathbf{v}_k'$, of the four trajectories belonging to a contributing set.

\begin{figure}[!t]
\centerline{
\includegraphics[width=4.2cm,angle=0]{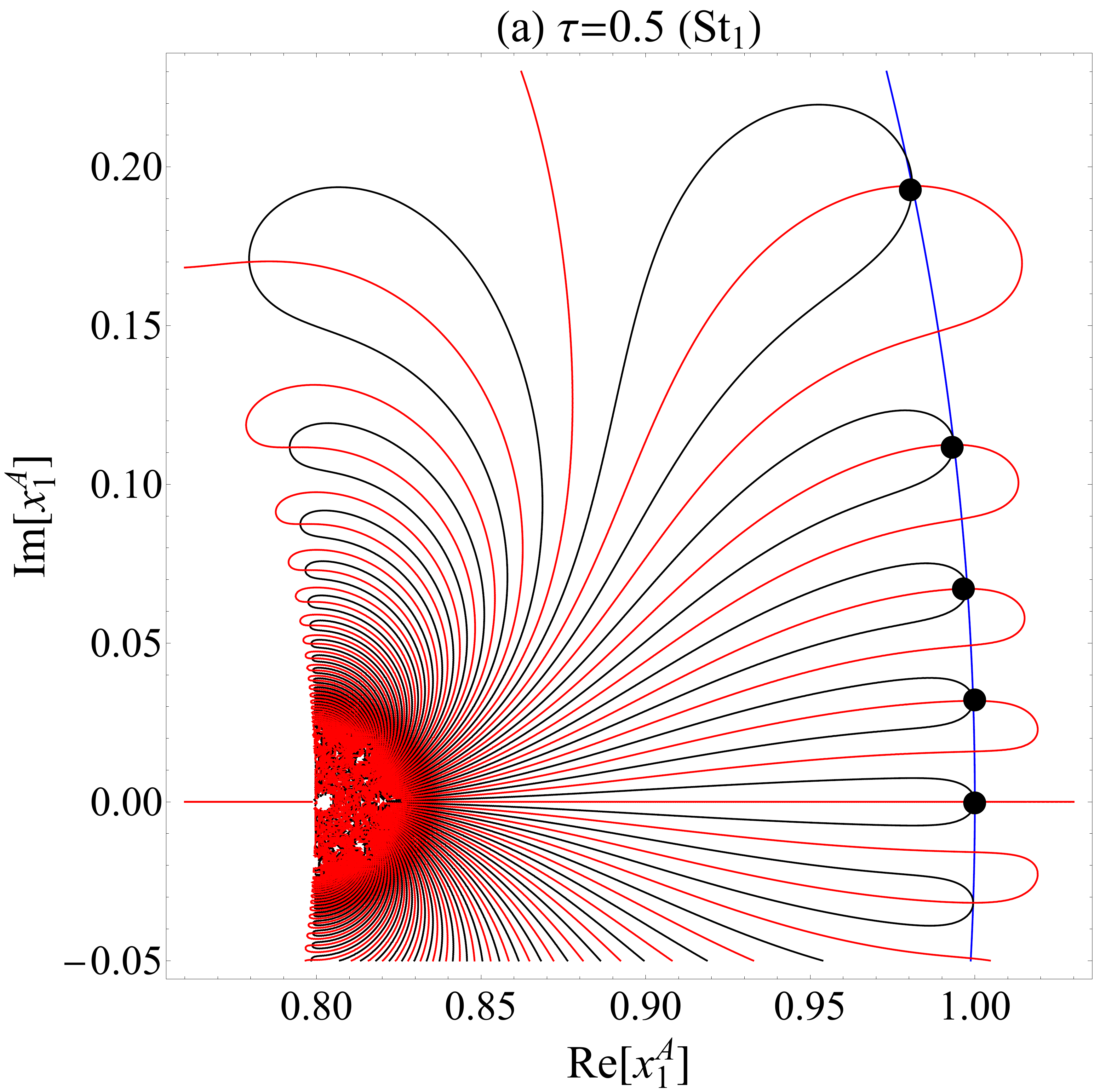}
\includegraphics[width=4.2cm,angle=0]{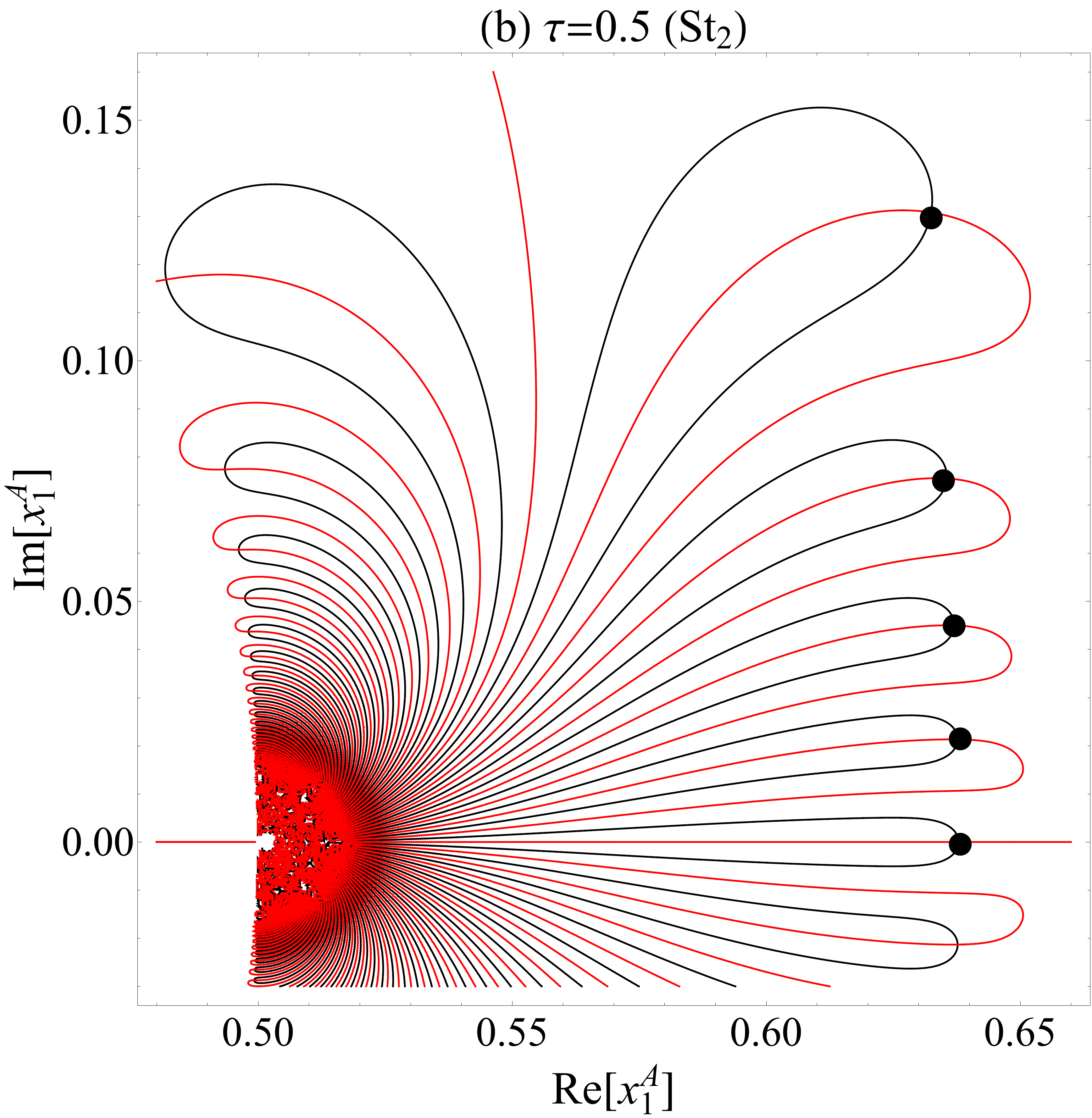}}
\vspace{0.2cm}
\centerline{
\includegraphics[width=4.2cm,angle=0]{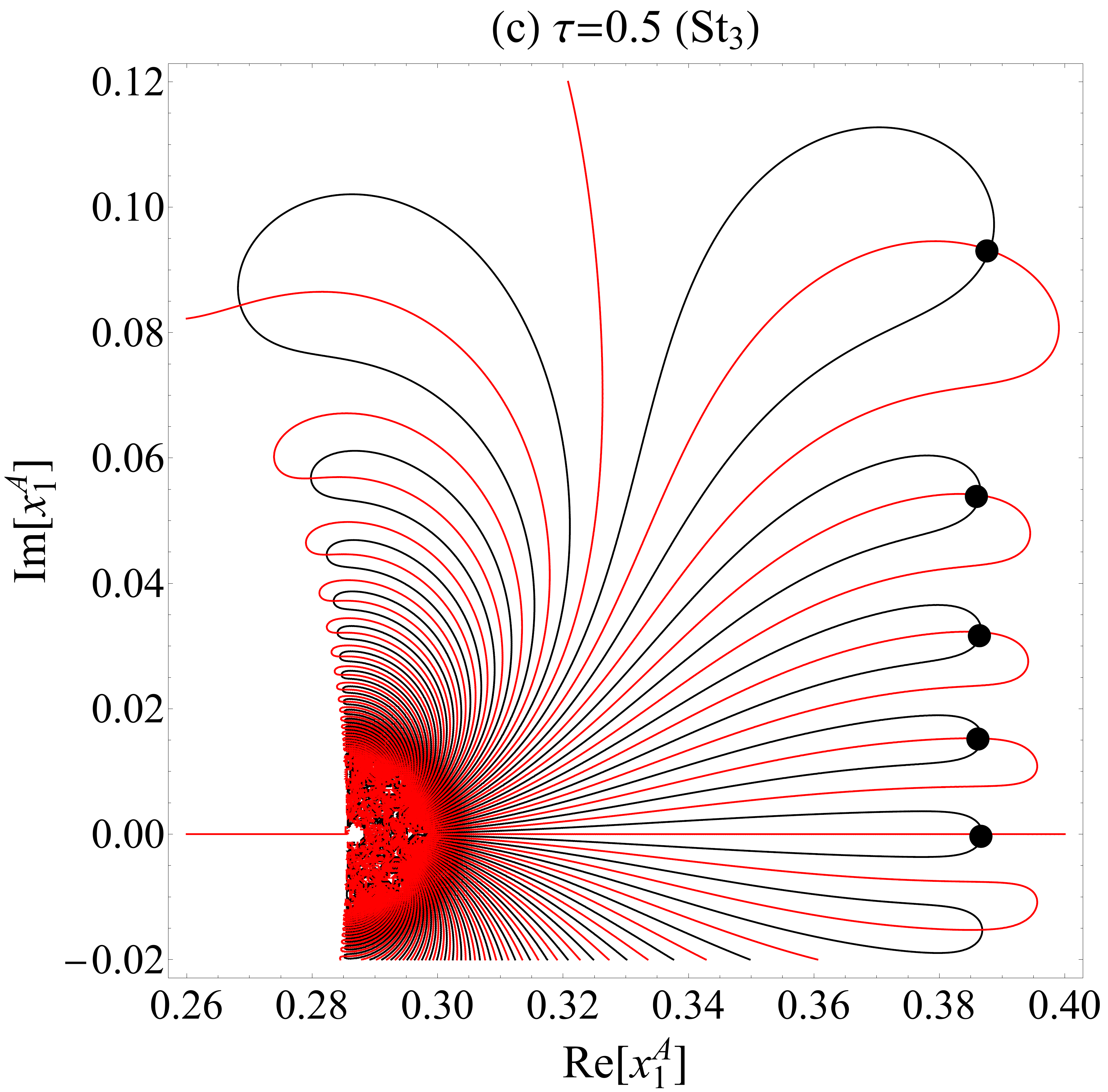}
\includegraphics[width=4.2cm,angle=0]{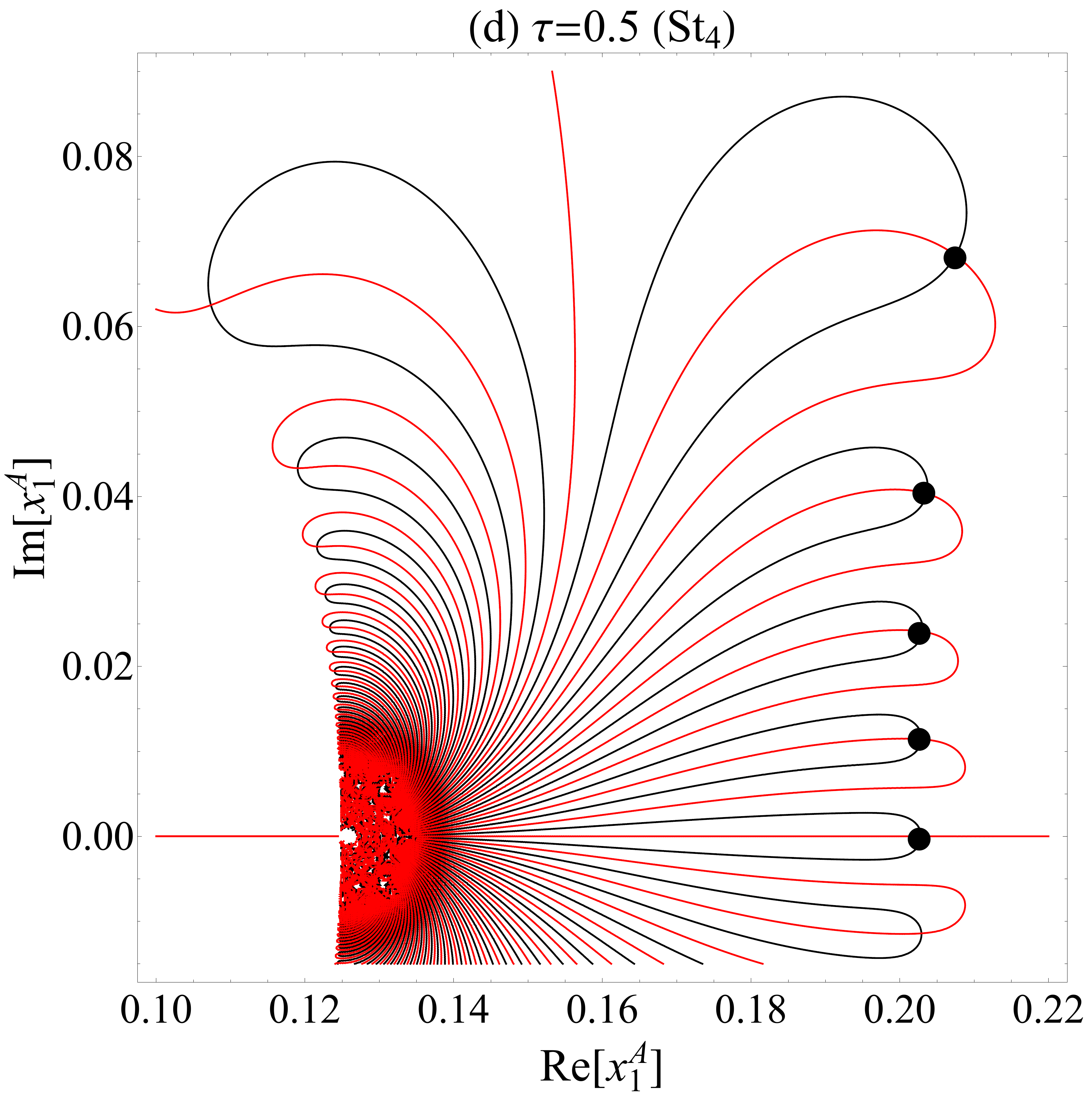}}
\caption{In panels (a)-(d), we show, respectively, the enlarged image of the structures $\mathrm{St}_1$,\ldots,$\mathrm{St}_4$, identified in Fig.~\ref{f2}(c). Small black circles are placed over some roots of $f(x_1^A)$ to indicate that they are considered in the calculation of Eq.~(\ref{Slinsemi}). }
\label{f3}
\end{figure}

\begin{figure}[!b]
\centerline{
\includegraphics[width=4.3cm,angle=0]{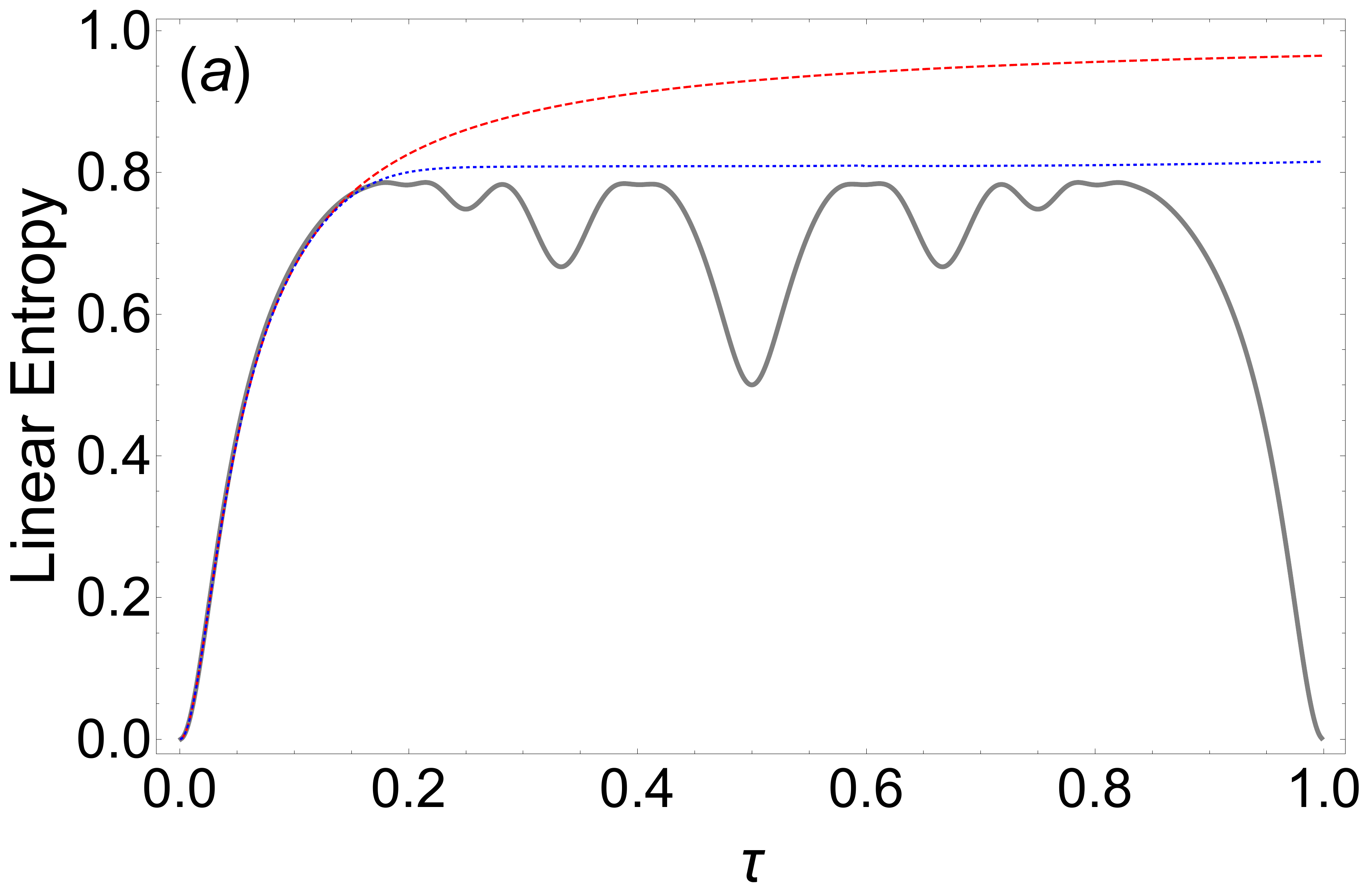}
\includegraphics[width=4.3cm,angle=0]{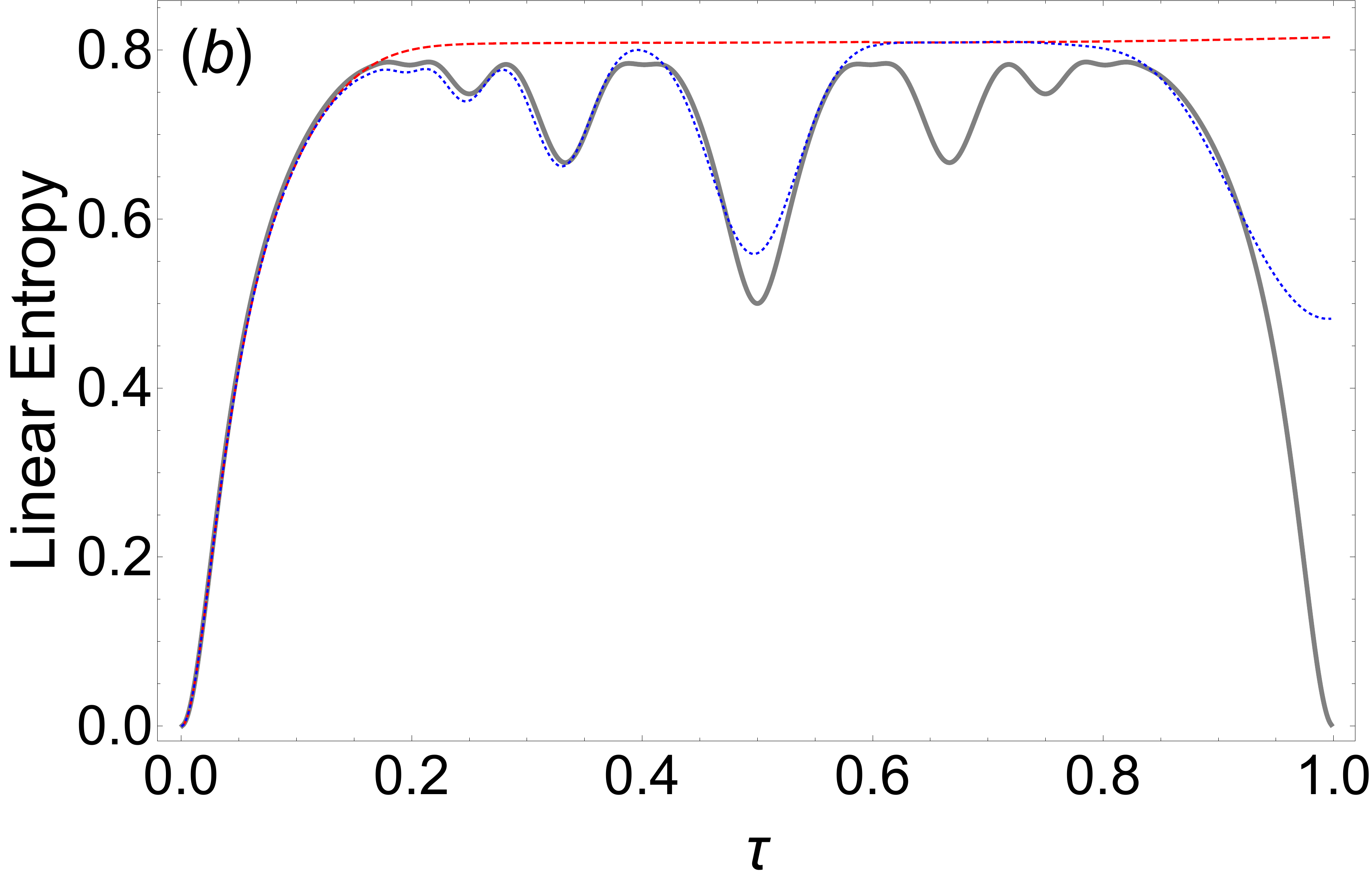}}
\centerline{
\includegraphics[width=4.3cm,angle=0]{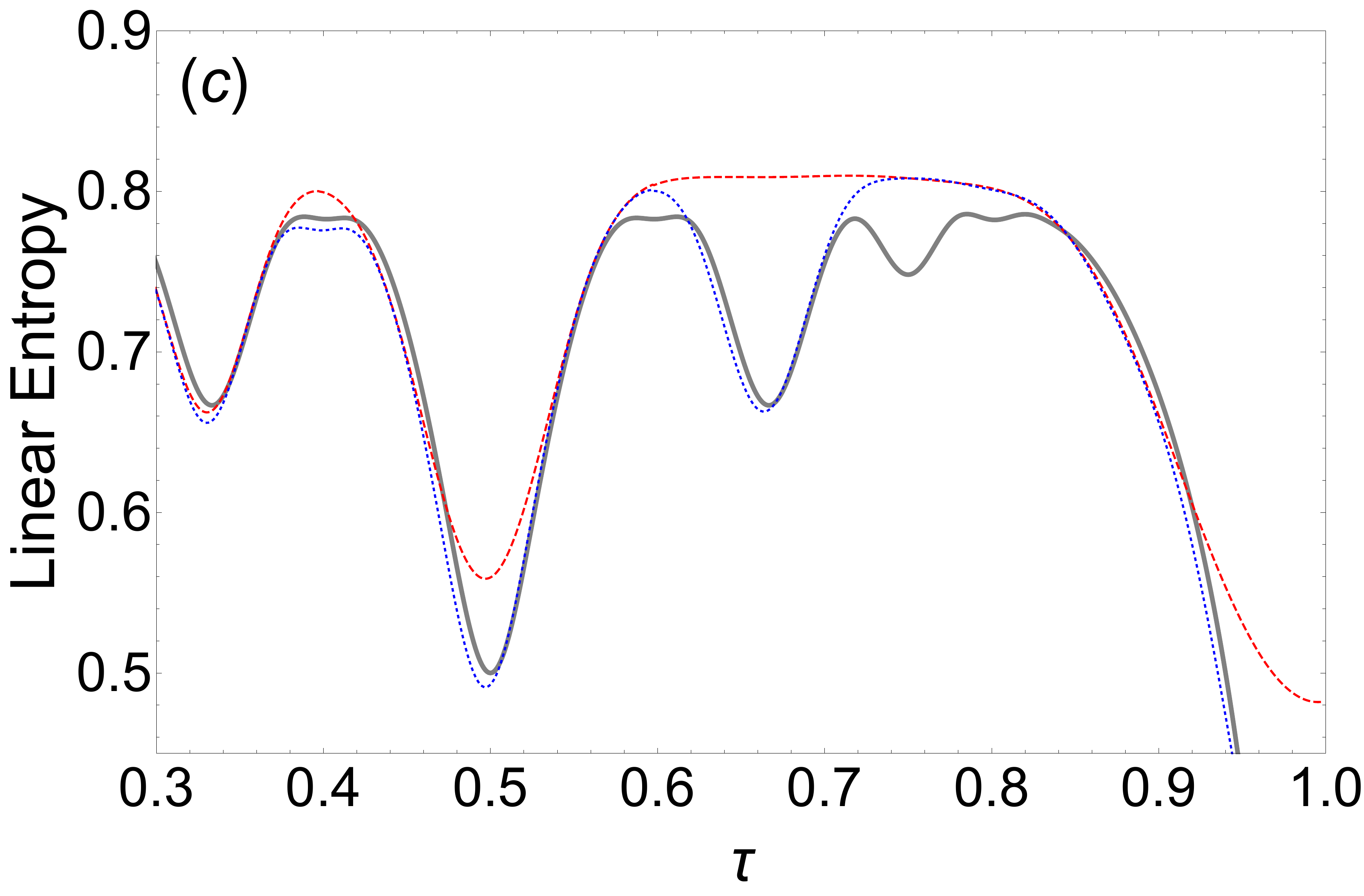}
\includegraphics[width=4.3cm,angle=0]{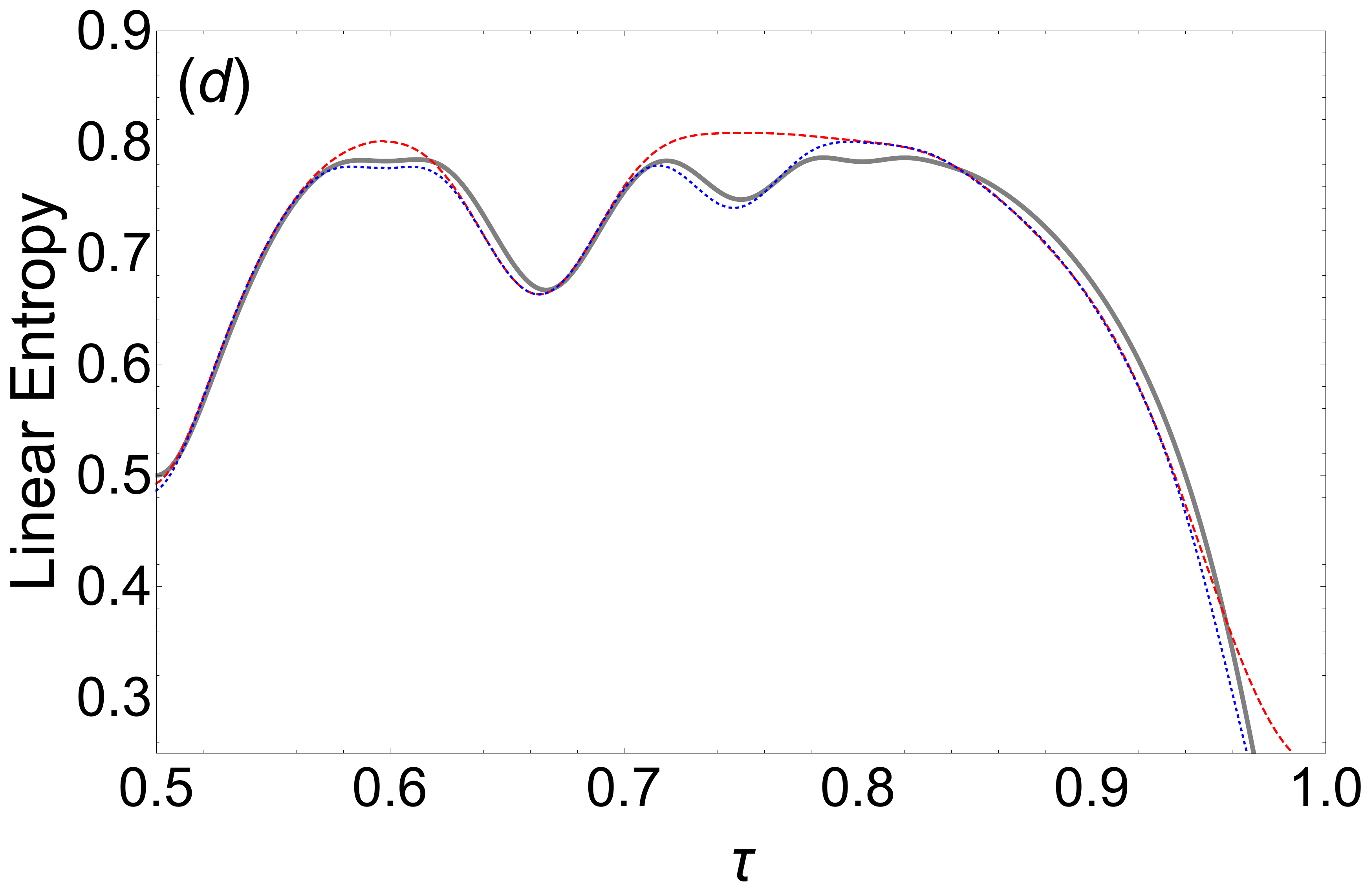}}
\caption{Linear entropy as a function of $\tau$. In panel (a), the red dashed curve represents the semiclassical entropy exclusively based on real trajectories, while the blue dotted line shows the result of Eq.~(\ref{Slinsemi}) when the roots marked in Fig.~\ref{f3}(a) are used. This result, for comparison purposes, is reproduced as the red dashed line in panel (b), where the blue dotted line represents $S_\mathrm{sc}$, improved by the inclusion of the roots marked in Fig.~\ref{f3}(b). The same logic is applied to panels (c) and (d), where the red dashed curve is copied from the previous plot, and the blue dotted curve shows the result of $S_{\mathrm{sc}}$ when the roots of Figs.~\ref{f3}(c) and~\ref{f3}(d), respectively, are included. The gray solid line appearing in all panels shows the quantum entropy~(\ref{SH2spins}). The numerical parameters used here are shown in Eq.~(\ref{nv}).}
\label{f4}
\end{figure}

In practice, to each solution of the transcendental equation, we assign a set of trajectories which, in principle, should be included in $S_\mathrm{sc}$. That is, for the present application, the task of finding contributing trajectories is equivalent to get solutions of~(\ref{eqtrans}). Then, given its importance, we now focus on some properties of this equation. First, we point out that $x_1^A=1$ is a solution, for any value of $T$. In this case, notice that all four trajectories have the same initial conditions $\mathbf{u}_k'=\mathbf{s}_0$ and $\mathbf{v}_k'=\mathbf{s}_0^*$, implying that they are real and identical. If we consider only this set of trajectories to evaluate Eq.~(\ref{Slinsemi}), as we already said, we get the same results as Ref.~\cite{pra2012}, which are illustrated in Fig.~\ref{f1} through the red dashed curve. Clearly, {\em real} trajectories provide a good approximation for the entropy~(\ref{SH2spins}), but only for the first stage of the time evolution. 

Extending the accuracy of the semiclassical entropy to longer values of time necessarily involves other solutions of Eq.~(\ref{eqtrans}). For $T=0$, however, it can be easily shown that the only solution is $x_1^A=1$. Fortunately, when $T$ increases, other solutions arise, part of them from the region around the origin while others arise from infinity. This behavior is illustrated in Figs.~\ref{f2}(a)-\ref{f2}(c), where we show some contours of $f(x_1^A)$ in the complex plane $x_1^A$. There, black and red curves refer to the $\mathrm{Re}[f(x_1^A)]=0$ and $\mathrm{Im}[f(x_1^A)]=0$, respectively. Intersection points of these two curves are, therefore, the roots of $f(x_1^A)$. In Fig.~\ref{f2}(a), built for a short value of time $\tau=0.01$, notice that only the solution $x_1^A=1$ appears. The others cannot be seen because either they are too close to the origin or too far from it, and their contribution to $S_{\mathrm{sc}}$ was numerically proven to be negligible. These observations justify the fact that real trajectories are enough to approach Eq.~(\ref{SH2spins}), as $\tau\to0$. In Figs.~\ref{f2}(a) and~\ref{f2}(b), for $\tau=0.1$ and 0.5, respectively, other roots of $f(x_1^A)$ start to appear in the plots, indicating that new sets of complex trajectories become important to calculate $S_\mathrm{sc}$.

\begin{figure*}[!t]
\centerline{
\includegraphics[width=5cm,angle=0]{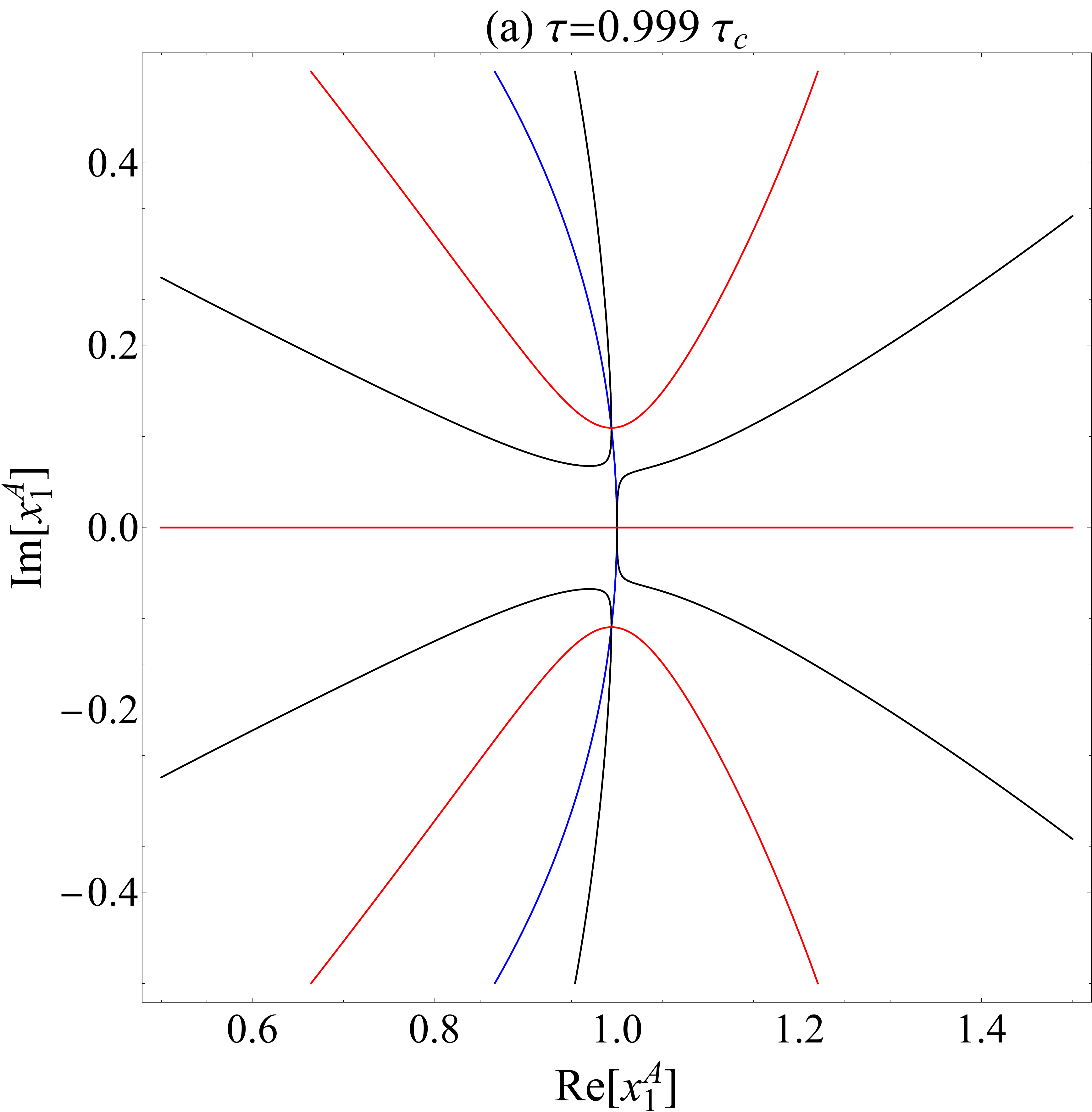}
\includegraphics[width=5cm,angle=0]{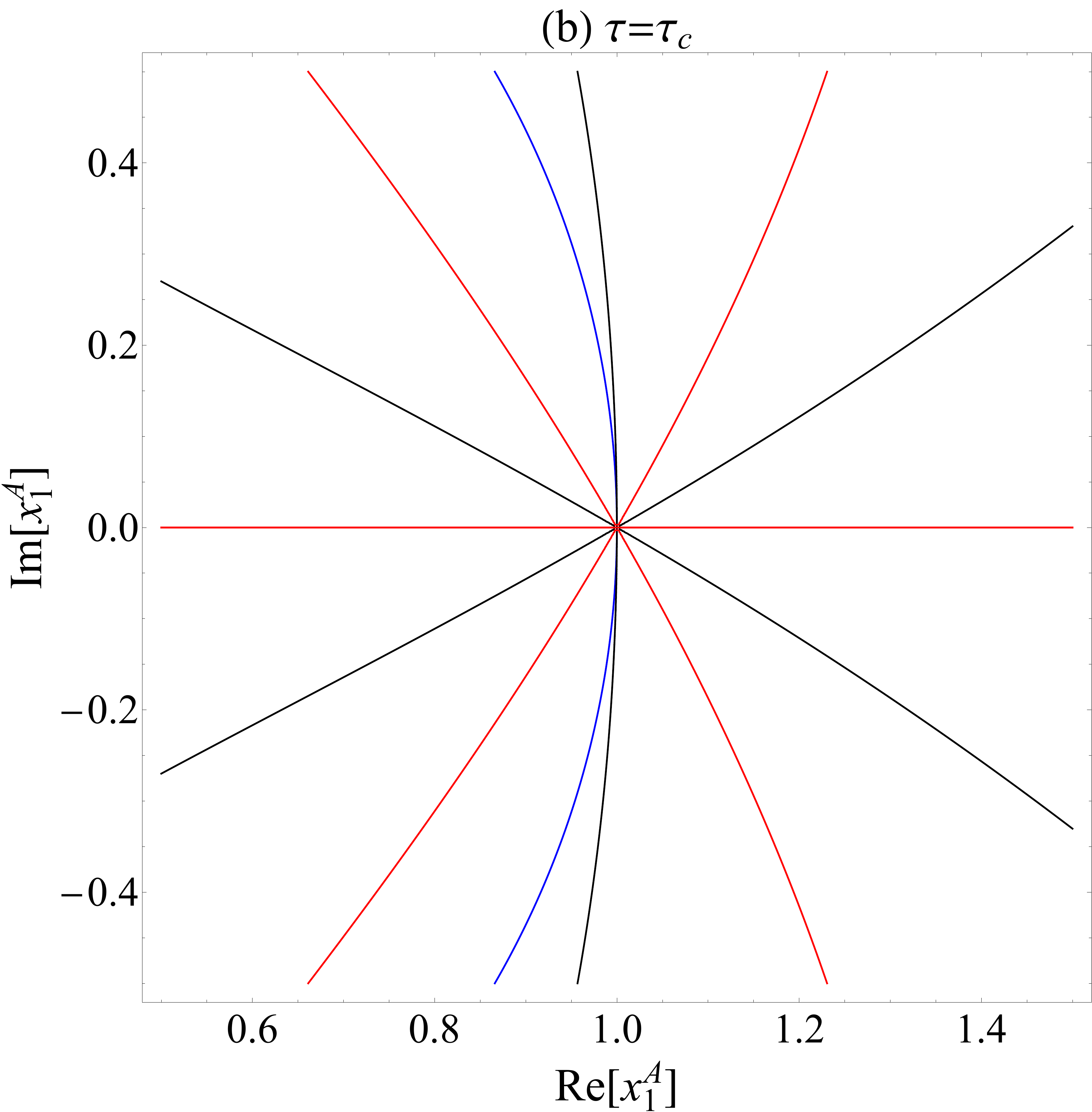}
\includegraphics[width=5cm,angle=0]{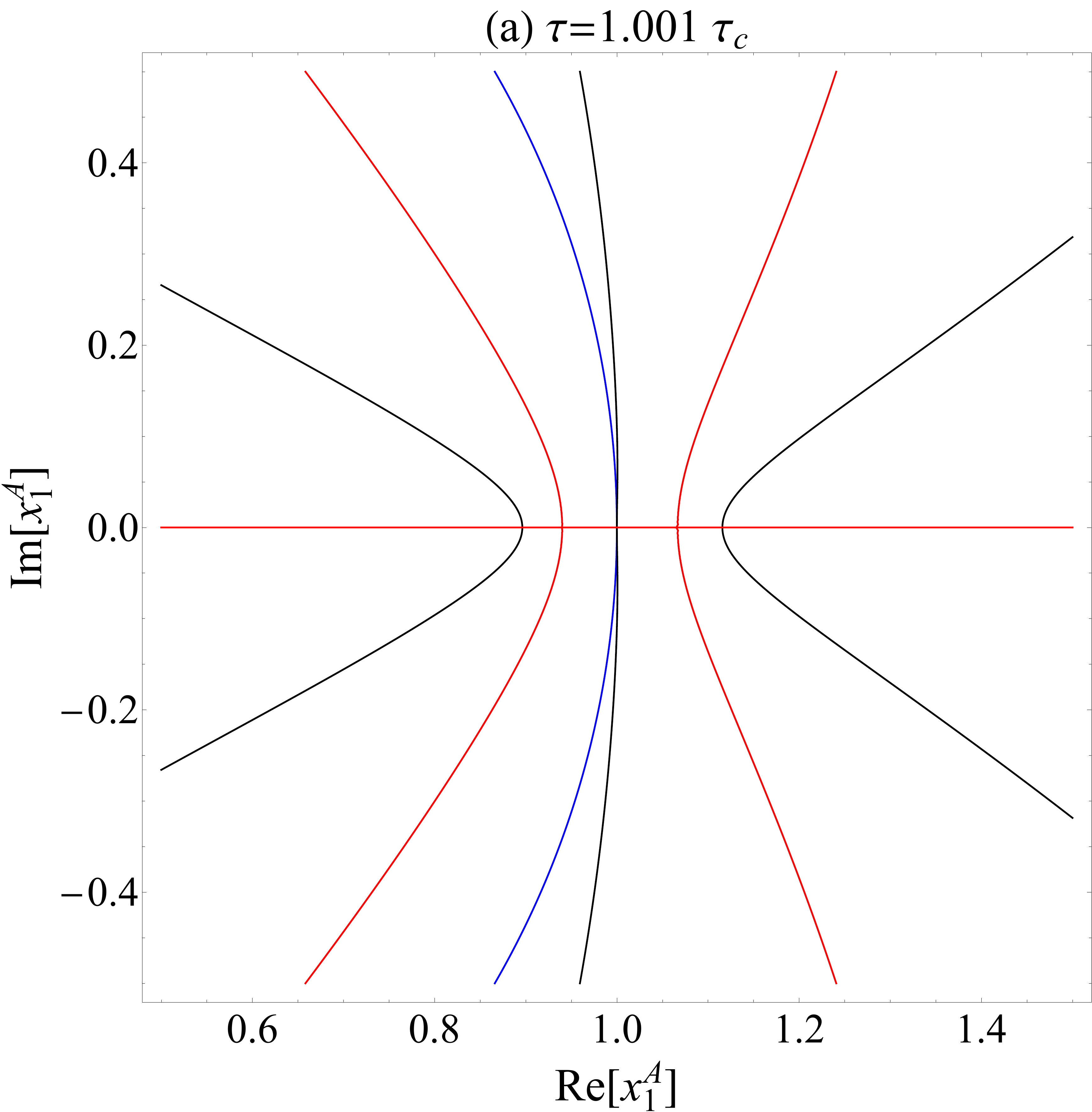}}
\caption{The same as Fig.~\ref{f2}, but with imaginary dimensionless time: (a) $\tau=0.999 \tau_c$, (b) $\tau=\tau_c$, and (c) $\tau=1.001 \tau_c$. }
\label{f5}
\end{figure*}

Before picking up each solution of the transcendental equation shown in the plots, it is important to systematize this procedure. We first notice that, for a given root $\bar{x}_1^A$, a simple inspection of Eq.~(\ref{eqtrans}) shows us that its complex conjugate $(\bar{x}_1^A)^*$ and their inverse, $1/\bar{x}_1^A$ and $1/(\bar{x}_1^A)^*$, are also roots of $f(x_1^A)$. Therefore, we only need to look for these solutions in the region inside the unitary circle, with $\mathrm{Im}[x_1^A] \ge 0$. Each observed root effectively represents four sets of four complex trajectories, and all of them should be considered, in principle. Of course, when a root lies exactly over the real axis or the unitary circle, this conclusion should be reconsidered because $\bar{x}_1^A=(\bar{x}_1^A)^*$, in the first case, and $1/\bar{x}_1^A=(\bar{x}_1^A)^*$, in the second.

Given these arguments, we first identify in Fig.~\ref{f2}(c) the structures named as $\mathrm{St}_1$, $\mathrm{St}_2$, \ldots, $\mathrm{St}_5$, and reproduce their magnified image (except for $\mathrm{St}_5$) in Figs.~\ref{f3}(a)-\ref{f3}(d). In each plot, we mark five roots of $f(x_1^A)$ with circles, in order to calculate $S_\mathrm{sc}$. Other intersection points seen in the search region were not considered because their contribution is negligible or unphysical. Of course, by varying $\tau$, all these points change. Therefore, to find the roots for all values of time, we recursively apply a proper routine based on the the Newton-Raphson method, to get a solution for $\tau+\delta\tau$, given that we know the root for~$\tau$.

Finally, in Fig.~\ref{f4}(a), we evaluate the contribution of the sets of complex trajectories indicated by the circles of Fig.~\ref{f3}(a). The gray solid line is the quantum result~$S_{\mathrm{pc}}$ [this curve also appears in Figs.~\ref{f3}(b)-\ref{f3}(d), for the sake of comparison], while the red dashed one is obtained from the inclusion of the real trajectories only. When the five new solutions are taken into account, the semiclassical approximation is clearly improved (blue dotted curve), but both the oscillatory behavior of~$S_{\mathrm{pc}}$ and its return to zero when $\tau\to1$ are not reproduced. In Fig.~\ref{f4}(b), new trajectories associated with the roots marked in $\mathrm{St}_2$ of Fig.~\ref{f3}(b) are considered, as represented by the blue dotted curve. Again, the inclusion of these new sets of complex trajectories substantially improves the semiclassical approximation for values of time until $\tau\approx 0.5$. Giving continuity, notice in Fig.~\ref{f4}(c) that the interval where quantum and semiclassical results (blue dotted curve) agree with each other increases to $\tau\approx 0.7$ when $\mathrm{St}_3$ of Fig.~\ref{f3}(c) is considered. Moreover, by using also the roots of $\mathrm{St}_4$ shown in Fig.~\ref{f3}(d), the accuracy seen in Fig.~\ref{f4}(d) becomes still better. At last, with the inclusion of the roots of $\mathrm{St}_5$, whose magnification is not shown in Fig.~\ref{f3}, all oscillatory behavior of~$S_{\mathrm{pc}}$ and also the return to zero at the end of the period are very satisfactorily reproduced, as shown by the blue dotted curve of Fig.~\ref{f1}.

With this example, we demonstrate that the semiclassical theory used to deduce $S_{\mathrm{sc}}$ can be quite successful. However, before finishing this section, we still have to develop the ideas presented in Sect.~\ref{ntv}. Here, we can calculate the determinant of the matrix~(\ref{Mstar}), finding
\begin{equation}
\det\bar{\mathbf{M}}^\star = 1 + 
\frac{16 j^2 |s_{0A}|^2|s_{0B}|^2 T^2}{\left(1+|s_{0A}|^2\right)^2\left(1+|s_{0B}|^2\right)^2}\neq0.
\label{detMstar}
\end{equation}
Therefore, we conclude that there is no set of complex trajectories arbitrarily close to the real one for any value of $T$. This result seems to be in contradiction to what is shown in Fig.~\ref{f3}(a), where there is a root of $f(x_1^A)$ very close to $x_1^A=1$, the point representing the real trajectory. In fact, Eq.~(\ref{detMstar}) means that these two points cannot coalesce, behavior that, numerically, we have really not found. Just to illustrate a mathematical situation where the sets of quasi-real trajectories exist, we will define the {\em complex} time 
\begin{equation}
T_{{c}}\equiv \pm i\frac{\left(1+|s_{0A}|^2\right)\left(1+|s_{0B}|^2\right)}{4j|s_{0A}||s_{0B}|},
\end{equation}
which amounts to the dimensionless $\tau_c \equiv T_c/T_{\mathrm{r}}$. For this value of time, we have $\det\bar{\mathbf{M}}^\star=0$, indicating the existence of complex contributing trajectories close to the real one. To check this conclusion, in Fig.~\ref{f5}, we plot the curves $\mathrm{Im}[f(x_1^A)]=0$ and $\mathrm{Re}[f(x_1^A)]=0$ for three complex values of $\tau$. In Fig.~\ref{f5}(a), for $\tau=0.999~\tau_c$, we see two roots of $f(x_1^A)$ over the unitary circle and very close to $x_1^A=1$. When $\tau=\tau_c$, the three solutions coalesce, as shown by Fig.~\ref{f5}(b). Right after, for $\tau=1.001\tau_c$, we verify other two roots over the real axis, moving away from the real contribution [Fig.~\ref{f5}(c)]. Finally, it is important to comment that the calculation presented in Sect.~\ref{ntv} have no influence in the calculation of $S_{\mathrm{sc}}$, showed here, but we decided to keep it in this work because it may be important in other applications. 

\section{Final remarks}
\label{fr}

Starting from the formula of the quantum linear entropy---given by Eqs.~(\ref{Sdef}) and~(\ref{Pscs})---, which contemplates a system of two spins initially prepared in a product of coherent states, we performed a semiclassical approximation resulting in Eq.~(\ref{Slinsemi})---the main product of the present paper. According to this approach, the entanglement dynamics between the spins is a function of sets of four mutually connected trajectories, originated from the equivalent classical description of the system. These {\em entangled-boundary-condition trajectories} live in an extended classical phase-space, obtained from the analytical continuation of the original one onto the complex domain. Concerning our previous works~\cite{pra2010,pra2012,arlans}, we point out that the present contribution effectively differs from Refs.~\cite{pra2010,arlans}, because we now deal with spin degrees of freedom, as well as Ref.~\cite{pra2012}, which only considers the ordinary real trajectories. As we see in Fig.~\ref{f1}, taking into account complex trajectories is incontestable to achieve excellent accuracy between quantum and semiclassical results. It is worth mentioning that, in similar semiclassical approaches, the inclusion of complex trajectories was already proven fundamental to mimic the quantum behavior~\cite{ribeiro05,marcel,ademir}.
 
We emphasize that the boundary conditions~(\ref{fbc}) and~(\ref{inibc}) characterizing the sets of contributing trajectories consists of potentially useful information. For example, it may help to clarify the questioning about the situation where, in opposition to common sense, regular classical dynamics is not directly associated with rapid entanglement growth~\cite{matzkin2006, matzkin2006EPL, matzkin2010, matzkin2011, pattanayak2017, ghose2019}. Moreover, scars of classical dynamics appearing in some plots of the quantum entropy~\cite{neil2016,pattanayak2017} may also find some explanation using the present results. We have to comment that these possible routes of investigation are still quite speculative, also because our approach possesses an additional difficulty to elucidate quantum-classical transition, which is understanding the connection between real and complex classical dynamics. Actually, this is the scenario that has motivated the study presented in Sect.~\ref{ntv}. 

At last, we report two straight future directions of our work. First, we intend to apply the theory to a quantum system whose classical counterpart is chaotic. The main difficulty, in this case, is the search for contributing trajectories. Due to the absence of analytical expressions for the dynamics, it is not possible to proceed as we did in Sect.~\ref{example1}, where this task was reduced to solving a system of equations. To fill this gap, a possible solution is to develop an algorithm to converge trial trajectories to those satisfying the boundary conditions~(\ref{fbc}) and~(\ref{inibc}), a strategy already used in a similar problem~\cite{ribeiro04}. We emphasize that a chaotic application would finally put this semiclassical theory in a position to be compared to many other works of the literature. The second direct continuation is testing Bell-type inequalities using our approach, in the same spirit as Ref.~\cite{silveira}. Given that such inequalities cannot be violated by any local classical theory, it seems that this is a fundamental test to theories pretending to imitate quantum mechanics.

\section*{Acknowledgements}

This study was financed by the brazilian agencies: Coordena\c{c}\~ao de Aperfei\c{c}oamento de Pessoal de N\'{i}vel Superior (CAPES), Conselho Nacional de Pesquisa (CNPq), and Instituto Nacional de Ci\^encia e Tecnologia/Informa\c{c}\~ao Qu\^antica (INCT-IQ, 465469/2014-0).


%


\begin{thebibliography}{99}

\bibitem{EPR} 
A. Einstein, B. Podolsky, and N. Rosen, Phys. Rev. {\bf 47}, 777 (1935).

\bibitem{Sch1935}
E. Schr\"{o}dinger, Math. Proc. Cambridge Philo. Soc. {\bf 31}, 555 (1935).

\bibitem{bell} 
J. S. Bell, Physics (N.Y.) {\bf 1}, 195 (1965).

\bibitem{amicoRMP}
L. Amico, R. Fazio, A. Osteroh, and V. Vedral, Rev. Mod. Phys. {\bf 80}, 517 (2008).

\bibitem{horodecki} 
R. Horodecki, P. Horodecki, M. Horodecki, and K. Horodecki, Rev. Mod. Phys. {\bf 81}, 865 (2009).

\bibitem{chuang} 
M. A. Nielsen and I. L. Chuang, {\em Quantum computation and quantum information} (Cambridge University Press, U.K., 2000).

\bibitem{kyoko} 
K.~Furuya, M.~C.~Nemes, and G.~Q.~Pellegrino, Phys. Rev. Lett. {\bf 25}, 5524 (1998).

\bibitem{sanders} 
S.~Ghose and B.~C.~Sanders, Phys. Rev. A {\bf 70}, 062315 (2004).

\bibitem{angelo04} 
R. M. Angelo, S. A. Vitiello, M. A. M. de Aguiar, and K. Furuya, Physica A {\bf 338}, 458 (2004).

\bibitem{escobar}
L.~F. Santos, G.~Rigolin, and C.~O.~Escobar, Phys. Rev. A {\bf 69}, 042304 (2004).

\bibitem{jacquod1} 
Ph.~Jacquod, Phys. Rev. Lett. {\bf 92}, 150403 (2004).

\bibitem{angelo05} 
R.~M.~Angelo and K.~Furuya, Phys. Rev. A {\bf 71}, 042321 (2005).

\bibitem{marcel2005}
M. Novaes, Ann. Phys. {\bf 318}, 308 (2005).

\bibitem{prosen05}
M. Znidaric and T. Prosen, Phys. Rev. A {\bf 71} 032103 (2005).

\bibitem{brumer2007} 
H. Han and P. Brumer, J. Phys. B {\bf 40}, S209 (2007).

\bibitem{jacquod2} 
Ph.~Jacquod and C.~Petitjean, Adv. Phys. {\bf 58}, 67 (2009).

\bibitem{pra2010} 
A.~D.~Ribeiro and R.~M.~Angelo, Phys. Rev. A {\bf 82}, 052335 (2010).

\bibitem{bonanca2011} 
M. V. S. Bonan\c{c}a, Phys. Rev. E {\bf 83}, 046214 (2011).

\bibitem{pra2012}
A. D. Ribeiro and R. M. Angelo, Phys. Rev. A {\bf 85}, 052312 (2012).

\bibitem{casati}
G.~Casati, I.~Guarneri, and J.~Reslen, Phys. Rev. E {\bf 85}, 036208 (2012).

\bibitem{berenstein}
C. T. Asplund and D. Berenstein, Ann. Phys. {\bf 366}, 113 (2016).

\bibitem{neil2016} 
C. Neil et al., Nat. Phys. {\bf 12}, 1037 (2016).

\bibitem{zheng2017}
F. Xu, C. C. Martens, and Y. Zheng, Phys. Rev. A {\bf 96}, 022138 (2017).

\bibitem{arlans}
A. J. S. Lara and A. D. Ribeiro, Phys. Rev. A {\bf 100}, 042123 (2019).

\bibitem{quach}
A. Piga, M. Lewenstein, and J. Q. Quach, Phys. Rev. A {\bf 99}, 032213 (2019).

\bibitem{pappalardi}
A. Lerose and S. Pappalardi, Phys. Rev. A {\bf 102}, 032404 (2020).

\bibitem{smerzi}
S.-C. Li, L Pezz\`{e}, and A. Smerzi, Phys. Rev. A {\bf 103}, 052417 (2021).

\bibitem{matzkin2006}
M. Lombardi and A. Matzkin, Phys. Rev. A {\bf 73}, 062335 (2006).

\bibitem{matzkin2006EPL}
M. Lombardi and A. Matzkin, Europhys. Lett. {\bf 74}, 771 (2006).

\bibitem{matzkin2010}
M. Lombardi and A. Matzkin, Laser Phys. {\bf 20}, 1215 (2010).

\bibitem{matzkin2011} 
M. Lombardi and A. Matzkin, Phys. Rev. E {\bf 83}, 016207 (2011).

\bibitem{pattanayak2017}
B. Ruebeck, J. Lin, and A. K. Pattanayak, Phys. Rev. E {\bf 95}, 062222 (2017).

\bibitem{ghose2019}
M. Kumari and S. Ghose, Phys. Rev. A {\bf 99}, 042311 (2019).

\bibitem{radcliffe} J.~M.~Radcliffe, J. Phys. A {\bf 4}, 313 (1971).

\bibitem{klauderb} 
J.~R.~Klauder and B.~S.~Skagerstan, {\em Coherent States. Applications in Physics and Mathematical Physics} (World Scientific, Singapore, 1985).

\bibitem{perelomov} 
A.~Perelomov, {\em Generalized Coherent States and their Applications} (Springer-Verlag, Berlim, 1986).

\bibitem{gilmore} 
W.~M.~Zhang, D.~H.~Feng, and R.~Gilmore, Rev. Mod. Phys. {\bf 62}, 867 (1990).

\bibitem{gazeau} J.~P.~Gazeau, {\em Coherent States in Quantum Physics} (Wiley-Vch, Weinheim, 2009).

\bibitem{ur} E.~Schr\"odinger, Proceedings of the Prussian Academy of Sciences XIX, pp. 296-303 (1930), in Erwin Schr\"odinger, Gesammelte Abhandlungen, Band 3, Verlag der \"Osterreichschen Akkademie der Wissenschaften, Wien, pp. 348-356 (1984).

\bibitem{scsp1} 
J.~R.~Klauder, Phys. Rev. D {\bf 19}, 2349 (1979).

\bibitem{scsp2} 
Y.~Weissman, J. Phys. A {\bf 16}, 2693 (1983).

\bibitem{scsp3} 
E.~A.~Kochetov, J. Phys. A {\bf 31}, 4473 (1998).

\bibitem{aguiar01} 
M.~Baranger, M.~A.~M. de Aguiar, F.~Keck, H.~J.~Korsch, and B.~Schellaas, J. Phys. A {\bf 34}, 7227 (2001).

\bibitem{ribeiro04} 
A.~D.~Ribeiro, M.~A.~M.~de Aguiar, and M.~Baranger, Phys. Rev. E {\bf 69}, 066204 (2004).

\bibitem{garg1} 
C.~Braun and A.~Garg, J. Math. Phys. {\bf 48}, 032104 (2007).

\bibitem{sscsp1}
H.~Kuratsuji and T.~Suzuki, J. Math. Phys. {\bf 21}, 472 (1979).

\bibitem{sscsp2}
H.~Solari, J. Math. Phys. {\bf 28}, 1097 (1987).

\bibitem{sscsp3}
V.~R.~Vieira and P.~D.~Sacramento, Nucl. Phys. B {\bf 448}, 331 (1995). 

\bibitem{sscsp4}
E.~A.~Kochetov, J. Math. Phys. {\bf 36}, 4667 (1995). 

\bibitem{sscsp5}
M.~Stone, K.~S.~Park and A.~Garg, J. Math. Phys. {\bf 41}, 8025 (2000). 

\bibitem{garg2} 
C.~Braun and A.~Garg, J. Math. Phys. {\bf 48}, 102104 (2007).

\bibitem{ribeiro06}
A.~D.~Ribeiro, M.~A.~M~de Aguiar and A.~F.~R.~de Toledo Piza, J. Phys. A {\bf 39}, 3085 (2006).

\bibitem{thiago}
T.~F.~Viscondi and M.~A.~M~de Aguiar, J. Math. Phys. {\bf 52}, 052104 (2011).

\bibitem{foggiatto} 
A. L. Foggiatto, R. M. Angelo, and A. D. Ribeiro, Prog. Theor. Exp. Phys. {\bf 2017}, 103A01 (2017).

\bibitem{bleistein} 
N.~Bleistein and R.~A.~Handelsman, {\em Asymptotic Expansion of Integrals} (Dover, New York, 1986).

\bibitem{ribeiro05}
M.~A.~M. de Aguiar, M.~Baranger, L.~Jaubert, F.~Parisio, and A. D.~Ribeiro, J. Phys. A: Math. Gen. {\bf 38}, 4645 (2005).

\bibitem{marcel}
M.~Novaes, Phys. Rev. A {\bf 72}, 042102 (2005).

\bibitem{ademir}
A. L. Xavier, Jr. and M. A. M. de Aguiar, Phys. Rev. Lett. {\bf 79}, 3323 (1997).

\bibitem{silveira}
L. S. Silveira and R. M. Angelo, Phys. Rev. A {\bf 95}, 062105 (2017).

\end{thebibliography}
\end{document}